\documentclass[prd,superscriptaddress,
preprint,
showpacs, preprintnumbers,
amsfonts,amsmath,amssymb]{revtex4}
\usepackage{epsf,psfrag}
\usepackage{graphicx}
\usepackage{dcolumn}
\usepackage{bm}
\catcode`\@=11
\def \be  {\begin{equation}}
\def \ee  {\end{equation}}
\def \ba  {\begin{eqnarray}}
\def \ea  {\end{eqnarray}}
\def \baa {\begin{eqnarray*}}
\def \eaa {\end{eqnarray*}}
\def \bb  {\begin {thebibliography} }
\def \eb  {\end{thebibliography}}
\def \lab #1 {\label{#1}}

\def \matrix #1 {\left(\begin{array}{cc} #1 \end{array}\right)}

\def \Im {\mathop{\rm Im}\nolimits}
\def \Re {\mathop{\rm Re}\nolimits}

\newcommand \vev [1] {\langle{#1}\rangle}

\newcommand{\as}{\ifmmode\alpha_{\rm s}\else{$\alpha_{\rm s}$}\fi}
\newcommand{\asbar}{\ifmmode\bar{\alpha}_{\rm s}\else{$\bar{\alpha}_{\rm s}$}\fi}

\font\cmss=cmss12 
\def\inbar{\,\vrule height1.5ex width.4pt depth0pt}
\def\IC{\relax\hbox{$\inbar\kern-.3em{\rm C}$}}
\def\IZ{\relax{\hbox{\cmss Z\kern-.4em Z}}}
\def\IR{{\hbox{{\rm I}\kern-.2em\hbox{\rm R}}}}

\def\IP{{\hbox{{\rm I}\kern-.2em\hbox{\rm P}}}}
\def\II{\hbox{{1}\kern-.25em\hbox{l}}}

\def\app{%
\def\theequation {\Alph{section}.\arabic{equation}}
}%
\def \theequation {\arabic{section}.\arabic{equation}}

\begin{document}

\preprint{RUB-TPII-10/05}

\title{Evolution Equation for Generalized Parton Distributions}

 \author{M. Kirch}
\affiliation{Institut f\"ur Theoretische Physik, Ruhr-Universit\"at Bochum,
                          D-44780 Bochum, Germany}
\author{A. Manashov}
\altaffiliation[Also at ]{Department of Theoretical Physics,  Sankt-Petersburg State University,
St.-Petersburg, Russia}
\affiliation{Institut f\"ur Theoretische Physik, Universit\"at
                         Regensburg, D-93040 Regensburg, Germany}

\author{A.~Sch\"afer}
\affiliation{Institut f\"ur Theoretische Physik, Universit\"at
                         Regensburg, D-93040 Regensburg, Germany}
\date{\today}

\begin{abstract}
The extension of the
method~\cite{MKS} for solving the leading order evolution equation
for Generalized Parton
Distributions (GPDs) is presented. We obtain
the solution of 
the evolution equation both
for the flavor nonsinglet quark GPD and singlet quark and gluon GPDs.
The properties of the solution and, in particular, the asymptotic form
of  GPDs in the small $x$ and $\xi$ region
are discussed. 
\end{abstract}
\maketitle

\section{Introduction}
\label{intr}
The outstanding problem in the theory of strong interactions
is to understand how  hadrons  are built from quarks and gluons.
The essential information on the internal structure of  hadrons can
be obtained from  hard scattering processes. In such processes
the duration of the interaction between a projectile and a fast moving hadron is
small and the latter can be resolved into separate quarks and gluons (partons).
Thus in hard scattering processes a hadron can be considered as a
collection
of non-interacting partons. This description  of hadrons emerged
from  the study of
deep inelastic scattering (DIS) and evolved into  what is nowadays known
as the parton model of hadrons
and found its  fundamental justification  in QCD.

The key ingredient of the parton model are the parton densities
which
encode information on the distribution of  partons with respect to  longitudinal
momentum.
The recent trend in this field is to study  processes
which can provide additional  information (e.g. distribution
of  partons in the transverse plane,  their angular momentum)
about the internal structure of hadrons.
Such information is encoded in so-called
Generalized Parton Distributions (GPDs)
which were introduced in Refs.~\cite{ditter,Ji96,Radyushkin96}
in the analysis  of Deeply Virtual Compton Scattering (DVCS).
The GPD formalism is rather universal and 
applicable for the analysis of
many different processes, ranging from
completely inclusive  to exclusive ones (for comprehensive reviews see
Refs.~\cite{Diehl,GPV,BR}).

Since the GPDs are essentially nonperturbative quantities, no
methods exist so far, which allow  to obtain them directly from QCD.
At present   the only realistic way to get GPDs is to try to
extract them
from relevant experimental data \cite{GPD-A1,GPD-S1,GPD-A4}
(which test only convolution of GPDs and specific kernels)
and lattice data \cite{L-H3,L-G4,L-H4,L-G5} (which only can provide moments of GPDs).
Probably only global fits  will be selective enough to really
determine more than the most dominant GPDs. Such fits naturally rely
heavily on $Q^2$-evolution. This problem can be already treated within
perturbative QCD.
The equation governing the
$Q^2$ dependence of GPDs -- the evolution equation --
is well known. Since the GPDs are related to  matrix elements
of certain operators, the evolution equation for GPDs follows from the
renormalization
group (RG) equation for the latter.
In  special cases the GPD evolution equation reduces to the famous
DGLAP~\cite{D,GL,AP} and ERBL \cite{ER,LB} evolution equations.
However,
the general structure of  solutions of the  evolution equation for GPDs
was poorly understood so far.

Several approaches~\cite{KM-1,Nor,SH-1,BGMS} were proposed in the
past to get a solution for
the  general evolution equation for GPDs.
All of them  try to
exploit the fact that Gegenbauer moments of GPDs have a simple scale dependence.
However, the  Gegenbauer polynomials do not form a complete set in the
region where the GPDs are defined. This leads to the known problems when one tries to
restore the GPDs by their Gegenbauer moments.
The numerical algorithms based on these methods
are rather cumbersome and ineffective (see for discussion Ref.~\cite{Diehl}).

Recently, we presented  a new method for solving the GPD evolution equation~\cite{MKS}.
It is based  on the correct incorporation of the symmetry properties of the evolution
equation and provides a clear physical picture for the evolution.
In Ref.~\cite{MKS} we considered the evolution equation for the gluon GPDs
related to the matrix element of the twist two gluon operator
 and for simplicity neglected the effects of the mixing with
quark-antiquark operators. In the present paper we give a detailed description of
this approach for the example of the  flavor nonsinglet quark GPD and present the solution
of the evolution equation for the flavor singlet GPD which takes into
account  mixing between quark
and gluon operators. We would like to mention that an 
 approach similar to ours was recently proposed in Ref.~\cite{MS}.
It will be briefly discussed in
Sect.~\ref{s-q-gpd}.

The paper is organized as follows. In the Sect.~\ref{bg} we recall the definitions of GPDs
and  the corresponding evolution kernels. In  Sect.~\ref{sp} the symmetry properties
of the evolution equations are discussed. In  Sect.~\ref{s-q-gpd}
we consider in detail
the evolution equation for the isovector quark GPD and construct its solution.
In  Sect.~\ref{singlet} we present the solution of the evolution equation for
the singlet quark and gluon GPDs.  Sect.~\ref{summary} contains the concluding remarks.
In the Appendix some useful formulae are collected.

\setcounter{equation}{0}
\section{Background}
\label{bg}
Throughout this paper we shall use the notations of Ref.~\cite{Diehl}.
The GPDs are usually defined in terms of  matrix elements
of certain
non-local light-cone operators. We shall consider the isovector  operator
\be\label{O-a}
{\mathcal O}^a(z_1,z_2)~=~\bar q(z_1 n)\gamma^+ \tau^a q(z_2 n)
\ee
and quark and gluon isosinglet operators
\begin{subequations}
\label{Osinglet}
\ba\label{O-q}
{\mathcal Q}(z_1,z_2)&=&\bar q(z_1 n)\gamma^+ q(z_2 n)\,,
\\[2mm]
\label{O-g}
{\mathcal G}(z_1,z_2)&=&
G^{+\mu,i}(z_1 n)G^{\phantom{\mu}+,i}_\mu(z_2n)
\,.
\ea
\end{subequations}
Here $n$ is the light like vector, $n^2=0$, $\gamma^+=n\cdot\gamma$ and $\tau^a$
($a=1,2,3$ ) are the Pauli matrices. In  Eq.~(\ref{O-q}) a summation over  flavor
indices is implied.
The real coordinates $z_1,z_2$ specify the position of the operator on the light cone.
As usual, we shall imply but do not display explicitly,
the Wilson lines between the fields at points
$z_1$ and $z_2$.  Taking matrix elements of these operators between hadron states
one obtains GPDs related to these operators. 
For example, the pion isovector quark GPD is defined as follows
\be\label{pion-gpd-isov}
{i\epsilon^{abc}}\, H(x,\xi,t) =
\int \frac{d z}{2\pi}\, e^{ix 2z P^+}\,
  \langle \pi^b(p')| {\mathcal O}^c(-z,z)
  |\pi^a(p) \rangle \,,
\ee
where the kinematical variables are~\cite{Diehl}
\be\label{definitions}
P=\frac{p+p'}{2},\ \ \ \xi=\frac{p^{+}-{p'}^{+}}{p^{+}+{p'}^{+}},\ \ \
t=(p-p')^2\,.
\ee
Henceforth we shall suppress the $t-$dependence of GPDs since it is 
irrelevant for the 
evolution. It is convenient to fix the normalization of the
vector $n$ by the condition 
$(Pn)=P^+=1$.
With such a normalization the matrix elements of the
quark and gluon operators, Eqs.~(\ref{O-a}), (\ref{O-q}) and
~(\ref{O-g}), respectively,
as well as the coordinates $z_1,z_2$ become dimensionless.

The isosinglet GPDs are defined in a  similar way
(see e.g.  Ref.~\cite{Diehl} for the definitions).  We shall consider
the nucleon GPDs
\begin{subequations}\label{Fqg}
\begin{align} \label{nucleon-quark}
\,{\mathcal  F}^{q}(x,\xi)&=
\int \frac{d z}{2\pi}\, e^{ix 2z }\,
  \langle p'|\,  {\mathcal Q}(-z,z)\,
  \,|p \rangle \,, \\
\label{nucleon-gluon}
\, {\mathcal F}^{g}(x,\xi)&=
\int \frac{d z}{\pi}\, e^{ix 2z}\,
  \langle p'|\,{\mathcal G}(-z,z)
  \,|p \rangle \,,
\end{align}
\end{subequations}
The functions ${\mathcal F}^{q,(g)}$ contain  two different Lorentz structures and determine
four scalar GPDs, $H^{q}, H^{g}$ and $E^{q}, E^{g}$. As the Lorentz structure of GPDs
is  irrelevant for the discussion of their evolution  we shall treat the functions
${\mathcal F}^{q,(g)}$  as scalar functions. The GPDs $H,{\mathcal F}^q,{\mathcal F}^g$
 have support
in the $x-$interval $[-1,1]$. The skewness parameter $\xi$
(Eq.~(\ref{definitions})) is also restricted
to the interval $[-1,1]$ by kinematic, (in DVCS processes $\xi>0$).
It is standard to distinguish two different kinematical
regions, $|\xi|<|x|$, the so-called DGLAP region, and $|\xi|>|x|$ -- the ERBL or central region.
In these regions
GPDs describe different physical processes. In the central region a GPD describes the
emission of a quark antiquark (gluon) pair from the initial hadron, while in the DGLAP region
it describes the emission and absorption of a quark or an antiquark.

\subsection{Evolution equations: Isovector case}
\label{ee}
All functions $H, {\mathcal F}^{q}, {\mathcal F}^{g}$ depend on the
normalization scale $\mu$ at which the operators~(\ref{O-a}),
(\ref{O-q}) and (\ref{O-g}) are defined.
This dependence is governed by the renormalization group (RG) equation for the operators
in question. We shall use the ``coordinate space'' version of this equation~\cite{BB}.
The RG equation for the operator ${\mathcal O}^a$ at one loop order takes form
\be\label{RGa}
\left(\mu\frac{\partial}{\partial\mu}+\beta(g)
\frac{\partial}{\partial g}\right) {\mathcal O}^a(z_1,z_2)=-
\frac{\alpha_s}{\pi}\,[\mathcal{H}\, {\mathcal O}^a](z_1,z_2)\,,
\ee
where 
$\alpha_s$ is the strong coupling constant and
the operator $\mathcal H$, which we shall refer to as Hamiltonian,
is an integral operator. Eq.~(\ref{RGa}) represents a concise
description of  renormalization of the local composite operators.
Indeed, expanding the l.h.s. and r.h.s of Eq.~(\ref{RGa}) in a Taylor series in $z_1,z_2$
one reproduces the RG equation for the local operators.
The Hamiltonian $\mathcal H$ (which encodes information on the mixing matrix)
can be found in Ref.~\cite{BB}. It is convenient to represent it in the following form
\ba\label{Ha}
{\mathcal H}&=&C_F\,{\mathbb H}\,,\\[2mm]
\label{Ha-1}
{\mathbb H}&=&{\mathbb H}_1^q-{\mathbb H}_{ 2}^q+{\mathbb I}/2\,,
\ea
where $C_F=(N_c^2-1)/2N_c$ and
the integral operators ${\mathbb H}_1^q,{\mathbb H}_2^q$ are defined as follows
\begin{subequations}\label{Hq}
\ba\label{H12}
\left[{\mathbb H}_1^q\varphi\right](z_1,z_2)&=&\int_0^1 d\alpha\frac{\bar\alpha}{\alpha}
\left(2\,{\varphi}(z_1,z_2)-{\varphi}(z_{12}^\alpha,z_2)-
{\varphi}(z_1,z_{21}^\alpha)
\right)\,,
\\[2mm]
\label{H2q}
\left[{\mathbb H}_{ 2}^q\varphi\right](z_1,z_2)&=&
\int_0^1 d\alpha\int_0^{\bar\alpha}d\beta\,
{\varphi}( z_{12}^{\alpha},z_{21}^\beta)\,.\ \ \ \ \ {}
\ea
\end{subequations}
We use here the standard notations $z_{ik}^\alpha=z_i\bar\alpha +z_k\alpha$,
$\bar\alpha =1-\alpha$. As will be discussed later, both  Hamiltonians
are  invariant under  collinear conformal transformations
and these two structures are the only structures which
can appear in the Hamiltonian at one loop level~(see
Refs.~\cite{BDM,BDKM}). Also we shall see
in which way these terms are responsible for the different evolution effects.

Let us now take the matrix element of both sides of
Eq.~(\ref{RGa}). Then we get
an equation for the function $\varphi^a(z_1,z_2)=\vev{out|{\mathcal O}^a(z_1,z_2)|in}$,
which is an ordinary function (not an operator) of two real variables $z_1,z_2$.
Obviously  it satisfies the same RG equation,~(\ref{RGa}), independently
of the choice of $in$ and $out$ states. Of course, the properties of the function
$\varphi^a(z_1,z_2)$ depend strongly on the choice of the $in$
and $out$ states.
Taking the matrix element between
vacuum state and one hadron state gives the hadronic wave function,
while taking the forward matrix element results in the parton
densities.
Usually one has to know all of these functions (wave functions,
usual and  generalized parton densities)
not in coordinate space but in momentum space.  In  momentum space
the RG equations look different for different physical situations. But it is useful to
remember that all of them  originate from one and the same RG equation
for composite operators.
In some sense the only difference between the DGLAP-, ERBL- and GPD-type evolution equation
lies in the initial conditions for the evolution. The functions which one wants to evolve
have different properties in all these cases.
Therefore we prefer to work in the coordinate representation and
only in the last step
perform the transformation to  momentum space.

For the isovector current we define the function $\varphi_\xi(z_1,z_2)$ by
\be\label{defH}
i\epsilon^{abc}\varphi_\xi(z_1,z_2)= \langle \pi^b(p')| {\mathcal O}^c(z_1,z_2)
  |\pi^a(p) \rangle \,.
\ee
It is related to the isovector GPD as follows
\begin{align}\label{phiH}
\varphi_\xi(z_1,z_2)&=2\,e^{-i\xi(z_1+z_2)}
\int dx\, e^{ix(z_1-z_2)} H(x,\xi)\,.
\end{align}
The subscript $\xi$ of the function $\varphi$ indicates that  this function has
 fixed total momentum. Under translation it transforms as
$\varphi_\xi(z_1+a,z_2+a)=e^{-2i\xi a}\varphi_\xi(z_1,z_2)$.
We also introduce a function given by the integral
\be\label{phiz}
\Phi_\xi(z)=\int dx \, e^{ix z} H(x,\xi)\,
\ee
Provided that the GPD is  a smooth function of $x$ the function $ \Phi_\xi(z)$
(and $\varphi_\xi(z_1,z_2)$)
vanishes fast
for $z=z_1-z_2\to \pm \infty$.
The function $\varphi_\xi(z_1,z_2)$, as a function of $z_1+z_2$,
is a plane wave. We shall present the solution of the evolution equation
for the function $H(x,\xi)$, however, in order to deal with  well defined
expressions on intermediate steps
it is convenient to regard the convolution of  $\varphi_\xi(z_1,z_2)$
with some smooth function $\nu(\xi)$,
\be\label{conv}
\varphi(z_1,z_2)=\int \frac{d\xi}{4\pi}\, \nu(\xi)\, \varphi_\xi(z_1,z_2)\,.
\ee
One can always assume that this new function $\varphi(z_1,z_2)$ vanishes sufficiently
fast as $z_1,z_2\to\pm\infty$.
Eq.~(\ref{conv}) is a Fourier transform with respect to the total
momentum $\xi$,
therefore the function $\varphi(z_1,z_2)$ satisfies the same
evolution equation~(\ref{RGa}).  We postpone its
solution till Sect.~\ref{s-q-gpd} and discuss first  the RG equation
for the singlet quark and gluon GPDs.

\subsection{Evolution equations: Singlet case}

The quark and gluon operators~(\ref{O-q}), (\ref{O-g}), mix under
renormalization.
However, the $C-$odd operator
\be\label{Codd}
{\mathcal Q}^{-}(z_1,z_2)={\mathcal Q}(z_1,z_2)+ {\mathcal Q}(z_2,z_1)
\ee
can not mix with gluons and evolves autonomously according to  Eq.~(\ref{RGa}),
contrary to  the $C-$even operator which we choose as
\be\label{Ceven}
{\mathcal Q}^{+}(z_1,z_2)=\frac{i}{2}\left({\mathcal Q}(z_1,z_2)-
{\mathcal Q}(z_2,z_1)\right)\,,
\ee
which mixes with the gluon operator~(\ref{O-g}). To write down the RG
equation in a compact form
we introduce the vector notation ${\mathcal O}=\{{\mathcal
O^q},{\mathcal O}^{g}\}$ where
${\mathcal O}^g={\mathcal G}$ and
${\mathcal O}^q={\mathcal Q}^{+}$. Then one gets
\be\label{RGG}
\left(\mu\frac{\partial}{\partial\mu}+\beta(g)
\frac{\partial}{\partial g}\right) {\mathcal O}^i(z_1,z_2)=-
\frac{\alpha_s}{\pi}\,[\mathcal{H}^{ik}\, {\mathcal O}^k](z_1,z_2)\,.
\ee
The Hamiltonians ${\mathcal H}^{ik}$ can be found in  Ref.~\cite{BB}.
Let us notice that we use here the conventional definition of the nonlocal gluon operator
which differs by a sign from that used
in Ref.~\cite{BB}. It results in a change of  sign for the
off-diagonal kernels.

The Hamiltonian ${\mathcal H}^{gg}$ has the following form
\be\label{Hgg}
{\mathcal H}^{gg}=N_c\left({\mathbb H}_1^g-{\mathbb H}_2^g\right)+
\left(\frac76N_c+\frac13n_f\right){\mathbb I}\,.
\ee
The integral operators ${\mathbb H}_1^g$, ${\mathbb H}_2^g$ are defined as follows
\begin{subequations}\label{Hg}
\ba\label{Hgg-12}
\left[{\mathbb H}_1^g\varphi\right](z_1,z_2)&=&\int_0^1 d\alpha\frac{\bar\alpha^2}{\alpha}
\left(2\,{\varphi}(z_1,z_2)-{\varphi}(z_{12}^\alpha,z_2)-
{\varphi}(z_1,z_{21}^\alpha)
\right)\,,\nonumber\\[2mm]
\label{H2g}\left[{\mathbb H}_{ 2}^g\varphi\right](z_1,z_2)&=&
\int_0^1 d\alpha\int_0^{\bar\alpha}d\beta\,\bar\alpha\,\bar\beta\, \omega\!\left(
\frac{\alpha\beta}{\bar\alpha\bar\beta}
\right)\,
{\varphi}( z_{12}^{\alpha},z_{21}^\beta)\,,\nonumber
\ea
\end{subequations}
where the weight function is $\omega(\tau)=4(1+2\tau)$.

The Hamiltonian $\mathcal{H}^{qq}$ coincides with
${\mathcal H}$,~(see Eqs.~(\ref{Ha}),(\ref{Ha-1}))
\ba\label{H-d}
\mathcal{H}^{qq}&=&{\mathcal H}\,.
\ea
The off-diagonal Hamiltonians read~\cite{BB}
\ba\label{H-12}
{\mathcal H}^{gq}&=&
C_F\, A^{-1}\,\left({\mathbb I}+2\,{\mathbb H}_2^q\right)\,,\\[2mm]
\label{H-21}
{\mathcal H}^{qg}&=&
n_f\, A\,{\mathbb H}_2^g\,,
\ea
where ${\mathbb H}_2^q$ is defined in Eq.~(\ref{H2q}) and the
Hamiltonian ${\mathbb H}_2^g$ has
the form (\ref{H2g}) with the weight function $\omega(\tau)=1+3\tau$. The operator
$A$ is the operator of multiplication by $(z_1-z_2)$
\be\label{A}
[A\varphi](z_1,z_2)=(z_1-z_2)\varphi(z_1,z_2)\,.
\ee
Further, in full analogy with the isovector case we introduce the functions
\be\label{Fz}
f_\xi^i(z_1,z_2)= \langle p'| {\mathcal O}^i(z_1,z_2)
  |p \rangle \,.
\ee
They are related to the nucleon GPDs~(\ref{nucleon-quark}),(\ref{nucleon-gluon}) as follows
\begin{subequations}\label{fGF}
\begin{align}
f_\xi^{g}(z_1,z_2)&=\,e^{-i\xi(z_1+z_2)}
\label{fG}
\int dx\, e^{ixz_{12}}\,{\mathcal F}^g(x,\xi)\,,\\[2mm]
f_\xi^{q}(z_1,z_2)&=
e^{-i\xi(z_1+z_2)}
\label{fQ}
\int dx\, e^{ixz_{12}}\,{\mathcal F}^{q,+}(x,\xi),
\end{align}
\end{subequations}
where $z_{12}=z_1-z_2$ and
\be\label{Fq+}
{\mathcal F}^{q,+}(x,\xi)=
{i}
\left({\mathcal F}^{q}(x,\xi)-{\mathcal F}^{q}(-x,\xi)\right)\,.
\ee
For later convenience we introduce the notations
\begin{subequations}\label{FF}
\begin{eqnarray}
{\mathcal F}_1(x,\xi)&\equiv&{\mathcal F}^{q}(x,\xi)-{\mathcal F}^{q}(-x,\xi)\,,\\
{\mathcal F}_2(x,\xi)&\equiv &{\mathcal F}^{g}(x,\xi)\,.
\end{eqnarray}
\end{subequations}
Convoluting $f_\xi^i$ with  a smooth function $\nu^i(\xi)$ one gets the function
\be\label{f}
f^i(z_1,z_2)=\int\frac{d\xi}{2\pi}\,\nu^i(\xi) f^i_\xi(z_1,z_2)
\ee
which vanishes for $z_1,z_2\to\pm\infty$.
Let us note also that the function $f^g(z_1,z_2)$ ($f^q(z_1,z_2)$) is symmetric
(antisymmetric) under  permutation of variables, $z_1\leftrightarrow z_2$
and this symmetry is preserved by the evolution.

\subsection{Local operators}

Let us discuss now the problems which arise in solving the
evolution equation for GPDs.
For the sake of simplicity we consider the isovector operator
${\mathcal O}^{a}(z_1,z_2)$. 
As we said above the nonlocal operator ${\mathcal O}^a(z_1,z_2)$
should be understood as the generating function for the local composite operators,
\be\label{loc}
{\mathcal O}^a(z_1,z_2)=\sum_{k,m=0}^\infty {z^k_1z_2^m} \,{\mathcal O}^a_{k,m}\,,
\ee
where ${\mathcal O}^a_{k,m}$ is $(nD)^k\bar q(0)\gamma^+ \tau^a (nD)^m q(0)/k!m!$
and $D_\mu$ is
the covariant derivative.
Inserting this expansion into Eq.~(\ref{RGa}) one gets an infinite set
of equations describing the renormalization of the local operators. Solving these equations
one finds a set of the  multiplicatively renormalized local operators of twist two.
They are
\be\label{local2}
\mathcal{O}^a_{N,k}(0) = (i\partial_+)^{N+k} \bar q(0)\,\gamma^+\tau^a
\mathrm{C}_N^{3/2}
\left( \frac{ \stackrel{\to}{D}_+ -
\stackrel{\gets}{D}_+ }{ \stackrel{\to}{\partial}_+ +
\stackrel{\gets}{\partial}_+} \right) q(0),
\ee
where
$\mathrm{C}_N^{3/2}$ is a Gegenbauer polynomial. These operators
enjoy  an autonomous scale dependence to one-loop order
\be\label{RGO}
\left(\mu\frac{\partial}{\partial\mu}+\beta(g)
\frac{\partial}{\partial g}\right) {\mathcal O}_{N,k}^a(0)=-
\frac{\alpha_s}{2\pi} \,\gamma_N\,{\mathcal O}_{N,k}^a(0)\,,
\ee
where $\gamma_N$ is the anomalous dimension of the operator.
One can rearrange the expansion~(\ref{loc}) in the following way~\cite{BB,BFLS}
\ba\label{rloc}
{\mathcal O}^a(z_1,z_2)&=&\sum_{N=0}^\infty B_N i^N (z_1-z_2)^N
\int_0^1\,du (u\bar u)^{N+1}\, {\mathcal O}_{N}^a(z_{12}^{u})\,.
\ea
Here $B_N=2(2N+3)/(N+1)!$ and
${\mathcal O}_{N}^{a}(z)\equiv{\mathcal O}_{N,0}^{a}(z)$.
Inserting the expansion (\ref{rloc}) into (\ref{defH}) one gets
\begin{align}
\label{Kiv}
\varphi_\xi(z_1,z_2)=& {\sqrt{2\pi}} e^{-i\xi(z_1+z_2)}
\sum_{N=0}^\infty i^{N}(2N+3)\, c_N(\xi)\, \Psi_N(z_{12})\,,
\end{align}
where $z_{12}=z_1-z_2$ and
\begin{subequations}\label{PcK}
\begin{eqnarray}
\label{def}
 \Psi_N(z_{12})&=&(\xi z_{12})^{-3/2} J_{N+3/2}(\xi z_{12})\,,\\[2mm]
\label{cN}
c_N(\xi)&=&\int_{-1}^{1}dx\, C_N^{3/2}\left(\frac{x}{\xi}\right) H(x,\xi)\,.
\end{eqnarray}
\end{subequations}
Here $J_\nu(z)$ is the Bessel function.
The coefficient $c_N(\xi)$ is related to the matrix element of the
local operator ${\mathcal O}^{c}_N$  as follows
\begin{equation}
\label{cNQ}
i\epsilon^{abc}\,\xi^Nc_N(\xi)=2^{-N-1} \langle \pi^b(p')| {\mathcal O}^c_N(0)
  |\pi^a(p) \rangle \,.
\end{equation}
Taking into account the tensor structure of  ${\mathcal O}^c_N(0)$
one concludes that the coefficient $\xi^Nc_N(\xi)$ is a
 polynomial in $\xi$
of degree $N$~\cite{Ji96}. This is  the so-called  polynomiality condition.
Since the operator ${\mathcal O}^c_N(0)$, Eq.~(\ref{local2}), is multiplicatively
renormalized its matrix element depends on
the scale in a simple manner
\begin{equation}\label{cNev}
c_N^{\mu_2}(\xi)= c_N^{\mu_1}(\xi)\, L^{-\gamma_N/b_0}\,,
\end{equation}
where $L=\alpha_s(\mu_1)/\alpha_s(\mu_2)$,
and        $b_0=\frac{11}{3}N_c-\frac23 n_f$.
Thus multiplying the coefficients $c_N(\xi)$
in Eq.~(\ref{Kiv}) by corresponding exponents one gets the solution of
the evolution equation~(\ref{RGa}).
The representation~(\ref{Kiv}) was obtained first in
Ref.~\cite{KM-1}.

The series in (\ref{Kiv}) converges for any $z_{12}$ but it converges non-uniformly
with respect to $z_{12}$ what causes some problems. Namely, in the Fourier transform
of (\ref{Kiv}) it is not possible to interchange the integration and summation.
Indeed, let us note that the Fourier transform of the function $\Psi_N(z)$
\begin{eqnarray}
i^N\int_{-\infty}^\infty dz e^{-ixz} \Psi_N(z)=
\frac{\sqrt{2\pi}\xi^{-1}}{(N+1)(N+2)}
\,\theta(x^2-\xi^2)
\left(1-\frac{x^2}{\xi^2}\right)\,C_N^{3/2}\left(\frac{x}{\xi}\right)
\end{eqnarray}
has  support in the central region $|x|\leq |\xi|$ only,
while the Fourier transform of $\varphi_\xi(z_1,z_2)$ is nonzero
in the whole $x-$region, $[-1,1]$.
In other words  taking the Fourier transform of (\ref{Kiv})
one obtains a formal representation for the GPD $H(x,\xi)$
\begin{equation}\label{formal}
H(x,\xi)=\frac1\xi\sum_{N=0}^\infty \omega_N\, c_N(\xi) 
\,\theta(x^2-\xi^2) \left(1-\frac{x^2}{\xi^2}\right)\,C_N^{3/2}\left(\frac{x}{\xi}\right)\,,
\end{equation}
where
$$
\omega_N=\frac12\frac{2N+3}{(N+1)(N+2)}\,,
$$
but this series  diverges if $H(x,\xi)$ is nonzero in the DGLAP region $|x|>\xi$.
Indeed, taking into account that for large $N$ and $t>1$, $C_N^{3/2}(t)\sim (t+\sqrt{t^2-1})^N$
one concludes that the coefficients $c_N(\xi)$ grow for large $N$ as  $\xi^{-N}$ and
one gets a divergence.

One can try to give sense to the series (\ref{formal}) by interpreting it as
a distribution, i.e. one
has at first to carry out a convolution with some test function and
then to
take the sum provided
that the latter converges. However, it is clear that the distribution defined in a such way,
$\widetilde H(x,\xi)$,  is different
from $H(x,\xi)$, since the former has support in the central region only.
\begin{figure}[t]
\psfrag{N}[cc][cc][1.2]{$ N$}
\psfrag{r}[cc][cc][1.2][90]{$\log|\bar r_n^N|$}
\centerline{\epsfxsize7.5cm\epsfbox{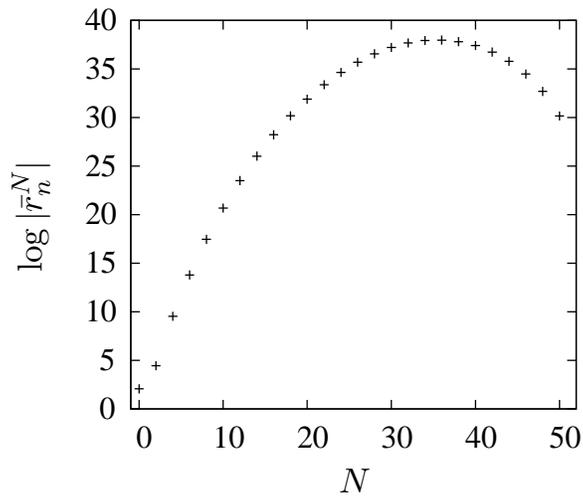}}
\caption[]{The coefficients $\log |{\bar r}^N_n|$  plotted as function of $N$ for $n=50$.}
\label{rNf}
\end{figure}
The problem  with the Fourier transform of  Eq.~(\ref{Kiv}) is related to the fact that
the Gegenbauer polynomials $C_N^{3/2}(x/\xi)$ are not orthogonal in the interval $[-1,1]$
provided $|\xi|<1$. Nevertheless they form a complete set and the function $H(x,\xi)$ can
be restored by its Gegenbauer moments $c_N(\xi)$, (see Eq.~(\ref{cN})).
The corresponding algorithm was suggested
in~Refs.~\cite{BGMS,MPW} and relies on the reexpansion of  some system of  orthogonal
polynomials on the interval $[-1,1]$, say Legendre polynomials, in terms of  Gegenbauer
polynomials,  $C_N^{3/2}(x/\xi)$,
$$
P_n(x)=\sum_{N=0}^n r^N_n(\xi)\, \xi^N\,C_N^{3/2}(x/\xi)\,.
$$
Obviously, one can also express the
Legendre moments in terms of the Gegenbauer moments
\begin{equation} \label{reex}
p_n(\xi)=\sum_{N=0}^n r^N_n(\xi) \,\xi^N c_N(\xi)\,.
\end{equation}
Then the series $\sum_{n=0}^\infty (n+1/2) p_n(\xi) P_n(x)$ converges
to the function $H(x,\xi)$.
Since the scale dependence of the coefficients $c_N(\xi)$ is
known (Eq.~(\ref{cNev}))
one can calculate the coefficients $p_n(\xi)$, and as a consequence the GPD $H(x,\xi)$,
at any scale. However, an attempt to apply this scheme in practice reveals the following
problem -- the coefficients $ r^N_n(\xi)$ grow fast with $n$. The explicit expression
for the coefficients $ r^N_n(\xi)$ can be found in \cite{BGMS,KM-1}.
For the sake of clarity  we consider the limiting case $\xi=0$.
In this limit one finds that $\xi^N c_N(\xi)\to 2^N \bar c_N$, where
$$
\bar c_N=\frac{2\Gamma(N+3/2)}{\sqrt{\pi}N!}\int_{-1}^1dx x^N H(x,0)\,.
$$
The remaining dependence of the coefficients $\bar c_N$ on $N$ is then weak. The coefficients
$r_N^n$ is nonzero only if $N$ and $n$ have the same parity, ( $n-N=2k$ ).
Denoting $\bar r_n^N=2^N\,r_N^n(0)$ one gets
$$
\bar r_n^N=\frac{(-1)^k(n+N)!}{2^nk!(n-k)!\Gamma(N+3/2)}\,.
$$
The logarithm of the coefficients $|\bar r_n^N|$ as function of $N$ for $n=50$ is
plotted in Fig.~\ref{rNf}.
One sees that  in the sum
$$
p_n=\sum_{N=0}^n \bar r_n^N \bar c_N
$$
each term is large while the sum itself should be a number of
order $O(1)$. Thus to calculate
the coefficients $p_n$ with a reasonable accuracy one has to calculate the coefficients $\bar c_N$
with ``infinite'' accuracy. It is clear that a numerical algorithm based on such a
reexpansion sooner or later will have an accuracy problem.
Thus, the methods for solving GPD evolution equations
suggested in Refs.~\cite{BGMS,MPW,KM-1}, although being mathematically correct,
result in inefficient numerical algorithms.
In what follows we suggest another method to solve the GPD
evolution equation
which relies heavily on the symmetry properties of the latter.

\setcounter{equation}{0}
\section{Symmetry properties}             \label{sp}
As it is well known the classical QCD Lagrangian is invariant under conformal transformations.
This symmetry, however, does not survive in the full
quantum theory due to the renormalization
effects. Nevertheless, at one-loop level the counterterms inherit the symmetry
of the classical Lagrangian~(for a review see e.g. Ref.~\cite{BKM03}).
This means that if two local (gauge invariant) operators are related  to each
other by a symmetry transformation,
${\mathcal O}^2=\delta_\epsilon{\mathcal O}^1$, their one loop
counterterms are related by the
same symmetry transformation, ${\Delta\mathcal O}^2=\delta_\epsilon{\Delta\mathcal O}^1$.
If one uses the formulation in terms of  nonlocal operators this
statement is translated
into
a statement about the invariance of the evolution kernel (Hamiltonian) with respect to
the symmetry (conformal) transformations.

We  consider  operators of a special type~---~the symmetric traceless operators of
twist two (Eqs.~(\ref{O-a}), (\ref{Osinglet})). This set of operators
is closed under  renormalization.  The restriction of
the  full conformal group  to this class of operators gives the
so-called collinear conformal group which is the ordinary $SL(2,R)$ group.
The generators of the $SL(2,R)$ group, $S^+,S^-,S^0$, satisfy the commutation
relations
\begin{equation}\label{sl2a}
[S^+,S^-]=2S^0,\ \ \ \ \ \ [S^0,S^\pm]=\pm S^\pm\,
\end{equation}
and can be chosen as
\begin{equation}\label{gensl2}
S^+=z^2\partial_z+2sz\,,\ \  S^-=-\partial_z\,,\ \ S^0=z\partial_z+s\,.
\end{equation}
Here the parameter $s$, the conformal spin, specifies the representation of
the $SL(2,R)$ group. Under an infinitesimal $SL(2,R)$ transformation a nonlocal
operator transforms as follows
\begin{equation}\label{inf}
\delta_\epsilon {\mathcal O}(z_1,z_2)=\epsilon_\alpha \left(
S_1^\alpha+S_2^\alpha
\right) {\mathcal O}(z_1,z_2)\,,
\end{equation}
where the spin operators should be taken in the representation corresponding
to the conformal spin $s=1$ for the quark operator (both singlet and nonsinglet),
Eqs.~(\ref{O-a}), (\ref{O-q}),
and to the conformal spin $s=3/2$ for the gluon operator, Eq.~(\ref{O-g}).
For later convenience we introduce the following notations for the two-particle
generators
\begin{subequations}\label{SS}
\begin{eqnarray}\label{SSq}
{S}_{q}^\alpha&=&({S}^\alpha_1+{S}^\alpha_2)|_{s=1}\,,\\
\label{SSg}
{S}_{g}^\alpha&=&({S}_1^\alpha+{S}_2^\alpha)|_{s=3/2}\,.
\end{eqnarray}
\end{subequations}
Since, as was discussed above, the conformal symmetry survives at one loop level
one gets
\begin{eqnarray}\label{SHq}
[{\mathcal H}, S_q^\alpha]=0
\end{eqnarray}
for the Hamiltonian~(\ref{Ha}). For the singlet case the Hamiltonian has the matrix form
(see Eqs.~(\ref{RGG})) therefore the symmetry relation takes form
\begin{eqnarray}\label{HHS}
\sum_{k={1,2}}[{\mathcal H}^{ik}, {\mathcal S}_{kj}^\alpha]=0\,,
\end{eqnarray}
where the matrix ${\mathcal S}$ is defined as
\begin{equation}\label{sgq}
 {\mathcal S}^\alpha=\left(\begin{array}{cc}S_g^\alpha & 0\\
                                             0    & S_q^\alpha\end{array}\right)\,.
\end{equation}
For the individual components Eq.~(\ref{HHS}) reads
\begin{subequations}\label{comp}
\begin{eqnarray}\label{comp-1}
 [{\mathcal H}^{gg}, S_g^\alpha]=0\,, && [{\mathcal H}^{qq}, S_q^\alpha]=0\,,\\[2mm]
\label{comp-2}
S_g^\alpha\,{\mathcal H}^{gq}={\mathcal H}^{gq}\,S_q^\alpha\,, &&
S_q^\alpha\,{\mathcal H}^{qg}={\mathcal H}^{qg}\,S_g^\alpha\,.
\end{eqnarray}
\end{subequations}
To solve the evolution equations it is helpful to take this symmetry into account.

Let us note that the transformations~(\ref{inf}) are the infinitesimal form of the
finite ($SL(2,R)$) transformations
\begin{equation}\label{finite}
{\mathcal O}(z_1,z_2)\to {(cz_1+d)^{-2s}(cz_2+d)^{-2s}}
{\mathcal O}(z'_1,z'_2)\,,
\end{equation}
where $z'_k=(az_k+b)/(cz_k+d)$, $ad-bc=1$, and the spin $s=1(3/2)$ for the quark (gluon) operators.
To proceed  we recall some facts about the representations of the $SL(2,R)$ group.

\subsection{$SL(2,R)$ group} \label{gsl2}
The $SL(2,R)$ group is the group of the real unimodular $2\times2$ matrices,
\begin{equation}
g=\left(\begin{array}{cc}a&b\\ c&d\end{array}\right),\ \ \ ab-cd=1\,.
\end{equation}
A detailed description of the representations of  $SL(2,R)$ group can be found
in \cite{Gelfand,Vilenkin}. Since the  $SL(2,R)$ group is a
noncompact group its unitary
representations are infinite-dimensional. They can be organized into three series, namely the
discrete series, the principal and supplementary continuous series~\cite{Gelfand}.
The last one will not appear in our analysis.

The representation of the unitary  principal continuous
series is determined by two parameters - the conformal
spin $s$, which takes values $s=1/2+i\rho$, where $\rho$ is
a real number; and by a discrete
parameter, the so-called signature
$\epsilon=0,1/2$.
 We need to consider the representations with signature $\epsilon=0$ only.
In this case, the representation, which is denoted by
$T^{\rho}$,
 can be realized by unitary operators
\begin{equation}\label{TT}
T(g^{-1})\psi(z)=\frac{1}{|cz+d|^{2s}}\psi(z'),\ \
z'=\frac{az+b}{cz+d}\,,
\end{equation}
acting on the Hilbert space of functions of a real variable
equipped with the scalar product
\begin{equation}\label{L2}
\vev{\psi_1|\psi_2}=\int_{-\infty}^\infty dz
\,\overline{\psi_1(z)}\,\psi_2(z)\,.
\end{equation}
The generators in this representation have the form~(\ref{gensl2}) with 
conformal spin  $s=1/2+i\rho$.

The representations of the discrete series are characterized by the integer or half-integer
conformal spin $s$, $s=1/2,1,3/2,\ldots$ and by the transformation law
\begin{equation}\label{TD}
T(g^{-1})\psi(z)=\frac{1}{(cz+d)^{2s}}\psi(z')\,.
\end{equation}
They fall into two classes usually denoted by $D_s^\pm$.
The difference between them lies in the space of  functions they are
defined on. The representations $D^+_s(D^-_s)$ are defined on the functions of the real
variable which can be continued analytically to the upper(lower) complex half-plane.
The scalar product in these cases reads
\begin{equation}\label{sd}
\vev{\psi_1|\psi_2}_\pm=\frac{2s-1}{\pi}\int_{\pm{\rm Im}z\geq 0} d^2z
(2{\rm Im} z)^{2s-2}
\overline{\psi_1(z)}\,\psi_2(z)\,.
\end{equation}
It is clear that functions from the representation space of $D^\pm_s$
can be represented in the form of a Fourier integral
\begin{equation}
\psi^\pm(x)=\int_0^\infty dp \,e^{\pm ipx}\,\psi^\pm(p)\,.
\end{equation}
In the momentum representation the scalar product reads
\begin{equation}\label{sdp}
\vev{\psi_1|\psi_2}_\pm=\Gamma(2s) \int_0^\infty dp p^{1-2s} \overline{\psi_1(p)}\,\psi_2(p)\,.
\end{equation}
One sees that in order to have a finite norm  $\psi(p)$ has to vanish
at $p=0$.

The tensor product of two unitary representations can  always be decomposed into
irreducible components. It will become clear later that we are interested in the tensor
product decomposition of the representations of the discrete series, $D^\pm_s$.
This decomposition depends strongly on whether one multiplies the representations
of the same type
or not. Namely, the decomposition of the tensor product  $D^+_s\otimes D^+_s$
(or $D^-_s\otimes D^-_s$)
contains only   representations of the same type
\begin{equation}\label{dpp}
D^\pm_s\otimes D^\pm_s=\sum_{n=0}^\infty \oplus D^\pm_{n+2s}\,.
\end{equation}
In contrast, the tensor product $D^+_s\otimes D^-_s$ (or $D^-_s\otimes D^+_s$)
can be decomposed into direct integral of the representations of the unitary principal
continuous series
\begin{equation}\label{dpc}
D^\pm_s\otimes D^\mp_s=\int_0^\infty d\rho \oplus T^\rho\,.
\end{equation}
\vskip 0.5cm

Now let us try to understand the group theoretical properties of GPDs in more
details. To be concrete we consider the quark GPD. In coordinate representation
the GPD $\varphi(z_1,z_2)$ inherits the symmetry properties of the operator
${\mathcal O}(z_1,z_2)$, Eq.~(\ref{finite}). Since the conformal spin $s=1$, this
transformation corresponds to the tensor product of the discrete series
representations of the $SL(2,R)$ group.
Let us represent the quark and antiquark fields by
\begin{eqnarray}\label{deq}
q(zn)&=&q^+(zn)+q^-(zn), \nonumber \\
\bar q(zn)&=&\bar q^+(zn)+\bar q^-(zn)\,,
\end{eqnarray}
where the components  $q^+$ and $\bar q^+$  contain the creation operators and thus, only
positive Fourier harmonics, $e^{i(pn)z}$, $p>0$, while the components
$q^-$ and $\bar q^-$  contain the annihilation operators
(negative Fourier harmonics, $e^{-i(pn)z}$, $p>0$).
Then, it is clear that e.g. the matrix element
$\vev{p'|\bar q^- (z_1n)\gamma^+ q^-(z_2 n)|p}$ as a function of two real variables, $z_1,z_2$,
admits an analytic continuation to the lower half-plane for each argument.
Therefore, the GPD $\varphi(z_1,z_2)$ can be represented as the sum of four functions
\begin{eqnarray}
\varphi(z_1,z_2)&=&\varphi^{++}(z_1,z_2)+\varphi^{+-}(z_1,z_2)+\varphi^{-+}(z_1,z_2)
+\varphi^{--}(z_1,z_2)\,,
\end{eqnarray}
where $\varphi^{\alpha\beta}\sim \vev{p'|\bar q^\alpha(z_1 n)\gamma^+ q^\beta(z_2 n)|p}$
and $\alpha(\beta)=\pm$. Each of the functions $\varphi^{\alpha\beta}(z_1,z_2)$ transforms
according to the tensor product of the representations $D^\alpha_s\otimes D^\beta_s$, with
$s=1$. We notice here that the functions $\varphi^{\pm\pm}(z_1,z_2)$ being transformed
to the momentum representation have support in the central region, while
the functions $\varphi^{\pm\mp}(z_1,z_2)$  in the DGLAP one.
Since the form of the tensor product decomposition
$D^\alpha\otimes D^\beta$
depends on whether  $\alpha=\beta$  or $\alpha\not=\beta$
 one has to decompose the functions
$\varphi^{\alpha\beta}$ over different set of functions in these two cases.

Let us note that our conclusion is in agreement with the solutions of the
evolution equation in the two limiting cases, $\xi=1$ and
$\xi=0$.
In the first one, $\xi=1$, when the DGLAP region shrinks to zero, to solve
the evolution equation one has to use the expansion in Gegenbauer polynomials,
in agreement with the decomposition~(\ref{dpp}). In the case of  forward scattering, $\xi=0$,
one should use a Mellin transform to solve the DGLAP evolution equation. This, in turn,
agrees  with the decomposition~(\ref{dpc}).

All this shows that the two kinematical regions,
"DGLAP" ($|x|>|\xi|$) and "central" ($|x|<|\xi|$) are quite
different, both from the physical
and mathematical point of view and, therefore, should be treated in a
different way.

\setcounter{equation}{0}
\section{Quark GPD}    \label{s-q-gpd}
In this section we consider in detail the case of the quark GPD~(\ref{pion-gpd-isov}).
We shall construct the solution of the evolution equation for  $H(x,\xi)$.
As it was explained earlier it is convenient to start analysis from the evolution
equation in the coordinate representation
\be\label{rgfa}
\left(\mu\frac{\partial}{\partial\mu}+\beta(g)
\frac{\partial}{\partial g}\right) \varphi(z_1,z_2)=-
\frac{\alpha_s}{\pi}\,[\mathcal{H}\, \varphi](z_1,z_2)\,,
\ee
where the Hamiltonian $\mathcal{H}$ and function $\varphi(z_1,z_2)$ are defined in
 Sect.~\ref{ee}.

We remind that the conventional strategy to solve   equations of this type is to expand
the function
$\varphi(z_1,z_2)$ in  eigenfunctions of the Hamiltonian ${\mathcal H}$,
$ \varphi(z_1,z_2)\sim\sum_E c_E\psi_E$. If  such an expansion is found,
the dependence of the expansion coefficients $c_E$ on the scale $\mu$
can be easily restored and,
provided that the corresponding sum can be effectively calculated, this determines
$\varphi$ at any scale $\mu$. We shall try to follow this scheme.

We start the discussion with the following remark. In order to talk rigorously about
expansion in eigenfunctions
of the Hamiltonian $\mathcal H$, one should specify the Hilbert space in which the
eigenvalue problem
for the operator has to be solved. In other words one has to define the
scalar product on
the space of functions $\varphi(z_1,z_2)$. It is clear  that this construction (scalar product,
Hilbert space) is external with respect to the problem. Indeed,
Eq.~(\ref{rgfa}) is an
integro-differential equation which knows nothing about the Hilbert space. However, to solve this
equation it is convenient to introduce the structure of  a Hilbert space, i.e. a scalar product,
and its choice is almost  completely in our hands.

The only condition which restricts the choice of the scalar product
is the requirement that  the function $\varphi(z_1,z_2)$   belongs to
the Hilbert space, i.e. it has to be normalizable with respect to the chosen scalar product.
This is the place where  physics imposes some constraints.
The other requirements  (such as the $SL(2,R)$ invariance in the case
under consideration)  are useful but not mandatory.

We choose the following $SL(2,R)$ invariant scalar product
\begin{equation}\label{sc}
\vev{\varphi_1|\varphi_2}=\int dz_1dz_2 (z_1-z_2)^2\,\overline{\varphi_1(z_1,z_2)}
\varphi_2(z_1,z_2)\,.
\end{equation}
Obviously, it is invariant under $SL(2,R)$ transformations
\begin{equation}\label{t-q}
{\varphi}_k(z_1,z_2)\to {(cz_1+d)^{-2s}(cz_2+d)^{-2s}}
{\varphi_k}(z'_1,z'_2)\,,
\end{equation}
with $s=1$ and $z'=(az+b)/(cz+d)$. Going to the momentum representation
\begin{equation}
\varphi(z_1,z_2)=\int \frac{d\xi dx}{2\pi}
 e^{i(-\xi(z_1+z_2)+x(z_1-z_2))}\varphi(x,\xi)
\end{equation}
one gets for the norm of the function $\varphi$
\begin{equation}\label{sc-p}
||\varphi||^2=\frac12\int d\xi dx\, |\partial_x \varphi(x,\xi)|^2\,.
\end{equation}
Then, taking into account Eqs.~(\ref{phiH}),~(\ref{conv}) one concludes that the physical GPD
has a finite norm~(\ref{sc-p})  if the integral $\int dx |\partial_x H(x,\xi)|^2$ is finite.
Since the GPD $H(x,\xi)$, as a function of $x$, has support on the
interval $[-1,1]$ this implies
that it should vanish faster than $(1\pm x)^{1/2}$ at the endpoints
and  be
continuous inside this interval. These requirements are in  agreement
with the
standard assumptions on the properties of quark GPDs~(see e.g. Refs.~\cite{Rad98,Yuan03}).

Let us remark  here that the other possible choice for the invariant scalar
product
\begin{equation}\label{sc-p-w}
||\varphi||^2\sim\int d\xi dx\, \frac{|\varphi(x,\xi)|^2}{|x^2-\xi^2|}\,,
\end{equation}
is ruled out.  Indeed, in this case the function $\varphi(x,\xi)$
in order to be normalizable
has to vanish at  $x=\pm \xi$, what is physically
unacceptable.

Next, it is useful to introduce a new function $\psi(z_1,z_2)$ related to $\varphi(z_1,z_2)$
by the simple relation
\begin{equation}\label{M}
\psi(z_1,z_2)=\left[A\varphi\right](z_1,z_2)=(z_1-z_2)\varphi(z_1,z_2)\,.
\end{equation}
One can easily check that if the function $\varphi$ transforms according to (\ref{t-q})
the function $\psi$ obeys the same transformation law with $s=1/2$. The map
$A:\varphi\to\psi$
is a one to one isometric map of the Hilbert space defined by the scalar product~(\ref{sc})
to the
standard,  ${\mathbf L}^2(R\times R)$ Hilbert space,
\begin{equation}\label{l2r}
\vev{\psi_1|\psi_2}=\int dz_1dz_2 \,\overline{\psi_1(z_1,z_2)}\,\psi_2(z_1,z_2)\,.
\end{equation}
The transformation of the Hilbert space
 ${\mathbf L}^2(R\times R)$
\begin{equation}\label{tpsi}
{\psi}(z_1,z_2)\to {(cz_1+d)^{-1}(cz_2+d)^{-1}}
{\psi}(z'_1,z'_2)\,,
\end{equation}
determines the unitary representation of the $SL(2,R)$ group
which is equivalent to the tensor product of the representations
\begin{equation}\label{Dhh}
{\mathcal D}^{\frac12,\frac12}=
\left(D^+_{1/2}\oplus D^-_{1/2}\right)\otimes \left(D^+_{1/2}\oplus D^-_{1/2}\right)\,.
\end{equation}
According to the Eqs.~(\ref{dpp}) and (\ref{dpc}) the decomposition of this representation
into  irreducible ones has form
\begin{equation}\label{DEC}
{\mathcal D}^{\frac12,\frac12}=
\sum_{n=0}^\infty \otimes D^\pm_{n+1}+2\int_0^\infty d\rho \otimes T^\rho\,.
\end{equation}

\subsection{Eigenvalue problem}\label{EvP}
We remind that our purpose is to construct the expansion of the GPD $\varphi(z_1,z_2)$
in terms of eigenfunctions of the Hamiltonian~(\ref{Ha}). Unfortunately, one can check that
the Hamiltonian $\mathcal H$ (Eq.~(\ref{Ha})) is non-hermitian with respect to
the scalar product~(\ref{sc}). It is possible  to find its eigenfunctions, (it will be done later)
but they do not form a complete set and, in general, it is not clear how to construct
the expansion over this set. Nevertheless such an expansion can be constructed. To do so
we use the following trick. As was explained before, the Hamiltonian
commutes with  the generators
of the $SL(2,R)$ transformations~(see Eq.~(\ref{SHq})), and as a consequence it commutes
with the Casimir operator of the $SL(2,R)$ group, $[{\mathcal H},{\mathbf J}^2]=0$,
\begin{equation}
{\mathbf J}^2=(\vec{S}_1+\vec{S}_2)^2=S_{12}^+S_{12}^-+S_{12}^0(S_{12}^0-I)\,,
\end{equation}
where $S_{12}^\alpha=S_1^\alpha+S_2^\alpha$ and $I$ is the identity
operator.
Taking into account the explicit form of the generators~(\ref{gensl2})
one finds that in  the  case $s_1=s_2=s$ the Casimir operator takes form
\begin{equation}\label{Cas}
{\mathbf J}^2=-z_{12}^{2-2s}\,\partial_1\partial_2 \,z_{12}^{2s}\,,
\end{equation}
where $z_{12}\equiv z_1-z_2$. For $s=1$ this operator is
a self-adjoint operator
with respect to the scalar product (\ref{sc}) (and for $s=1/2$ with respect to the
scalar product~(\ref{l2r})). So one can expand any function over the set of
eigenfunctions of the Casimir operator. Nevertheless, the operators ${\mathcal H}$
and ${\mathbf J}^2$, despite the fact that they commute, have
different eigenfunctions.
However, one of the crucial points of our approach is that, as
we shall show,  the expansion in eigenfunctions of
the Casimir operator
can be transformed into an expansion in  eigenfunctions of
the Hamiltonian.
\vskip 0.5cm

So as a first step we  find the eigenfunctions of the Casimir operator
${\mathbf J}^2$~(\ref{Cas}) for the conformal spin $s=1/2$. This problem is
equivalent to the decomposition of the representation of the $SL(2,R)$ group
into irreducible ones and its solution is well known~\cite{Pk}.
The Casimir operator has  eigenfunctions of  both the discrete and continuous spectrum.
The eigenvalues of the Casimir operators are usually written in the form $j(j-1)$.
The parameter $j$ is called conformal spin and takes integer values for
the eigenfunctions of the discrete spectrum
and is $j=1/2+i\rho$, with $\rho$ being real and positive,
for the eigenfunctions of the continuous spectrum.

For our purpose it is convenient to choose the eigenfunctions in the form
$\psi(z_1,z_2)\sim e^{-i\xi(z_1+z_2)} \Psi(z_{12})$.  Inserting this function
into the equation
\begin{equation}
{\mathbf J}^2\, \psi(z_1,z_2)=j(j-1)\psi(z_1,z_2)\,
\end{equation}
 one finds that the function $z_{12}^{1/2}\Psi(z_{12})$ satisfies the equation for Bessel
functions, $J_{\pm (j-1/2)}(|\xi|z_{12})$.
It is therefore convenient to define the function
\begin{equation}\label{psij}
\Psi_{j}^\xi(z)=
e^{-i\frac{\pi}{2}(j-1/2)}\,z^{-1/2} \,J_{j-1/2}\left(|\xi|z\right)\,.
\end{equation}
For integer $j$ it is a single valued function of $z$ in the whole complex plane.
For non-integer $j$ it has a cut from $0$ to $-\infty$.

The eigenfunctions of the discrete spectrum, $P^\xi_j(z_1,z_2)$, $j=n+1$, have the form
\begin{equation}\label{Pdj}
P^\xi_j(z_1,z_2)= e^{-i\xi(z_1+z_2)}\Psi_{j}^\xi(z_{12})\,.
\end{equation}
Calculating the scalar product one finds
\begin{equation}\label{scdd}
\vev{j',\xi'|j,\xi}=2\pi\delta(\xi-\xi')\,\frac{\delta_{jj'}}{2j-1}\,,
\end{equation}
where we used the standard notation for the scalar product of $P^{\xi}_j$ and  $P^{\xi'}_{j'}$.
The subspace spanned by the function $P^\xi_j(z_1,z_2)$, $\xi>0$ ($\xi<0$) corresponds to
the subspace $D^-_j$ ($D^+_j$) in the tensor product decomposition~(\ref{DEC}).

The eigenfunctions of the continuous spectrum, $P^{\xi,\pm}_j(z_1,z_2)$,
$j=1/2+i\rho$, (there are two eigenfunctions corresponding
to each value of $j$) have the form
\begin{equation}\label{Pcj}
P^{\xi,\pm}_j(z_1,z_2)= e^{-i\xi(z_1+z_2)}\Psi_{j}^{\xi,\pm}(z_{12})\,.
\end{equation}
The functions $\Psi_{j}^{\xi,\pm}(z_{12})$ are defined as follows
\begin{equation}\label{psipm}
\Psi_{j}^{\xi,\pm}(z)=
\frac1{2\cos\pi j}\left[\Psi_{j}^\xi(z_\pm)
-\Psi_{1-j}^\xi(z_\pm)\right ]\,,
\end{equation}
where $z_\pm=\pm z+i0$. Again, the subspaces spanned by the eigenfunctions
$P^{\xi,\pm}_j(z_1,z_2)$
correspond to two copies of the  subspaces $T^{\rho}$ in the tensor
product decomposition~(\ref{DEC}).
We notice also that these two functions differ only by
permutation of their arguments,
$$P^{\xi,+}_j(z_1,z_2)=P^{\xi,-}_j(z_2,z_1).$$
They are also invariant under the interchange $j\to 1-j$, $P^{\xi,+}_{j}(z_1,z_2)=
P^{\xi,+}_{1-j}(z_1,z_2)$.
Taking into account that the function $\Psi_{j}^{\xi,+}(z)$ is proportional
to a Hankel function
$$
\Psi_{j}^{\xi,+}(z)\sim (z+i0)^{-1/2} H^{(1)}_{j-1/2}(|\xi|(z+i0))
$$
one can check that $P^{\xi,+}_j(z_1,z_2)$ is an analytic function
of $z_1$
in the upper and of $z_2$ in the lower complex half-plane.
Next we conclude that from the properties of  Bessel functions follows that
the  functions $\Psi_{j}^{\xi,\pm}(z)$, as functions of $j$,
are  entire functions in the whole complex plane. For a fixed $j$
they vanish like $|z|^{-1}$ for  $|z|\to\infty$.

Calculating the scalar product of the eigenfunctions of the continuous spectrum one gets
\begin{eqnarray}
\vev{j',\xi',\alpha'|j,\xi,\alpha}&=&\pi \delta_{\alpha\alpha'}\delta(\xi-\xi')
[\delta(\rho-\rho')+\delta(\rho+\rho')]\frac{1}{\rho}\coth\pi\rho
\,,
\end{eqnarray}
where $\alpha(\alpha')=\pm$ and $j=1/2+i\rho$.

The eigenfunctions~(\ref{Pdj}),~(\ref{Pcj}) of the Casimir operator ${\mathbf J}^2$ form
a complete basis of the Hilbert space ${\mathbf L}^2(R\times R)$. Thus any function from
 ${\mathbf L}^2(R\times R)$ can be expanded as follows
\begin{align}\label{Exp12}
\psi(z_1,z_2)=
\int_{-\infty}^{\infty} \frac{d\xi}{2\pi}\left\{
\sum_{j=1}^\infty \omega(j) a_\xi(j) P_j^{\xi}
(z_1,z_2)
-i
\int_{\frac12-i\infty}^{\frac12+i\infty} dj
\omega^c(j) a^\pm_\xi(j) P_j^{\xi,\pm}(z_1,z_2)
\right\}\,.
\end{align}
Here $\omega(j)=2j-1$, $\omega^c(j)=(j-1/2)\cot\pi j$ and the expansion coefficients
$a_\xi(j)$ and $a_\xi^\pm(j)$ are given by the scalar products
\begin{eqnarray}\label{exc}
a_\xi(j)=\vev{j,\xi|\psi}\,,\ \ \ \
a_\xi^\pm(j)=\vev{j,\xi,\pm|\psi}\,.
\end{eqnarray}
It follows from the properties of the eigenfunctions  that the
coefficients
$a_\xi^\pm(j)$ are  entire functions of the conformal spin  $j$ in  the whole
complex plane satisfying the symmetry relation $a_\xi^\pm(j)=a_\xi^\pm(1-j)$.

Using the explicit expressions for the eigenfunctions~(\ref{Pdj}), (\ref{Pcj})
one can represent the expansion coefficients~(\ref{exc}) in the form
\begin{eqnarray}\label{exc-p}
a_\xi(j)&=&\varkappa (-1)^{j-1}\,
\int_{-|\xi|}^{|\xi|}dx\, P_{j-1}\left(\frac{x}{|\xi|}\right)\,\psi(x,\xi)\,,
\nonumber\\
a_\xi^\pm(j)&=&\varkappa\int_{|\xi|}^1 dx \,P_{j-1}\left(\frac{x}{|\xi|}\right)\,
\psi(\pm x,\xi)\,,
\end{eqnarray}
where $P_j(x)$ are the Legendre functions and
\begin{eqnarray}\label{coeff}
\varkappa&=&\frac12 e^{i\pi/4}\sqrt{{2\pi}/{|\xi|}}\,.
\end{eqnarray}
$\psi(x,\xi)$ is the Fourier
transform of $\psi(z_1,z_2)$ as in~(\ref{phiH}).
It can be shown that if the function $\psi(x,\xi)$  behaves like $(1-x)^{\alpha}$ for $x\to 1$,
then the coefficients $a_\xi^\pm(j)$ vanish as $j^{-3/2-\alpha}$
when $j$ goes to infinity along
the imaginary axis.

\begin{figure}[t]
\psfrag{i}[cc][cc][1.2]{$\Im j$}
\psfrag{r}[cc][cc][1.2]{$\Re j$}
\psfrag{c}[cc][cc][1.2]{$C_a$}
\psfrag{1}[cc][cc]{$1$}
\psfrag{2}[cc][cc]{$2$}
\centerline{\epsfxsize5cm\epsfbox{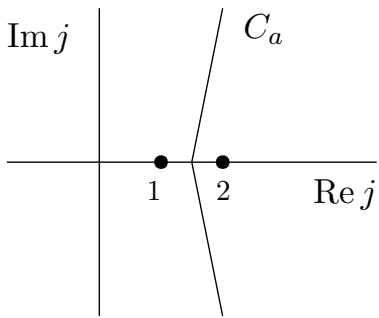}}
\caption[]{The integration contour $C_a$ in  Eq.~(\ref{Eh-B}).}
\label{cont}
\end{figure}

In the following we get rid of the Fourier integral over $\xi$ and construct the expansion
for the function $\psi_\xi(z)$ which is defined by
\begin{equation}
\psi(z_1,z_2)=\int \frac{d\xi}{2\pi} e^{-i\xi(z_1+z_2)}\,\psi_\xi(z_1-z_2)\,.
\end{equation}
Taking into account the symmetry property of the expansion coefficients,
$a_\xi^\pm(j)=a_\xi^\pm(1-j)$, and using Eq.~(\ref{psipm})
one derives
\begin{align}\label{Eh}
\psi_\xi(z)=  \sum_{j=1}^\infty \omega(j)\, a_\xi(j)\, \Psi_j^{\xi}(z)-
\frac{i}{2}\int_{\frac12-i\infty}^{\frac12+i\infty} \frac{dj}{\sin\pi j}
\, \omega(j)\,a^\pm_\xi(j)\, \Psi_j^{\xi}(z_\pm)\,.
\end{align}
Let us discuss the properties of the expansion~(\ref{Eh}). First, we note that
when $j$ goes to infinity along the imaginary axis the functions
$\Psi_j^{\xi}(z_\pm)$
behave as
\begin{equation}\label{asP}
\Psi_j^{\xi}(z_\pm) \sim 2^{-1/2}\frac{e^{-i\frac{\pi}2(j-1/2)}}{\Gamma(j+1/2)}
\left(\frac{z_\pm}{2}\right)^{j-1} \left(1+{\mathcal O}(1/j)\right)
\end{equation}
so that the integrals in (\ref{Eh}) converge. Moreover, shifting  the integration
contour to
the right we further improve the convergency of the integrals.
Second, let us remind that the function $\psi$ under consideration has the form~(\ref{M}),
$\psi_\xi(z)=z\varphi_\xi(z)$, and therefore vanishes at $z=0$.
For small $z$, $\Psi_j^\xi(z)\sim z^{j-1}$, thus all terms in the sum except the first one
with $j=1$ vanish as $z\to 0$. The integral over $j$ follows the line $\Re j=1/2$. In order
to study its small $z$ behavior let us shift the contour of
 integration to the line $\Re j=3/2$.
Since the integrand has a pole at $j=1$ due to $\sin\pi j$ one
should evaluate the residue
at this point.
Adding the contributions from residues and the first term in the sum one gets
$$
\left[a_\xi(1)+a_\xi^+(1)+a_\xi^-(1)\right]\Psi_{j=1}^\xi(z)\,.
$$
The term in the square brackets is zero, what is easy to see using Eq.(\ref{exc-p}) and
taking into account that $\psi_\xi(x)=i\partial_x\varphi_\xi(x)$.  Thus
one gets the following expansion for the function $\varphi_\xi(z)=\psi_\xi(z)/z$
\begin{align}\label{Eh-1}
\varphi_\xi(z)=z^{-1}~\left\{\sum_{j=2}^\infty\, \omega(j)\, a_\xi(j)\,
\Psi_j^{\xi}(z)~-~
\frac{i}{2}\int_C
\frac{dj}{\sin\pi j}\,
 {\omega(j)}\,a^\pm_\xi(j)\, \Psi_j^{\xi}(z_\pm)\right\}\,,
\end{align}
where the integration follows the line $ \Re j=k$, $1<k<2$.
We shall show below that this is an expansion in eigenfunctions of the Hamiltonian.
But before we want to notice that due to the asymptotics (\ref{asP}) the  
integration contour  in (\ref{Eh-1}) can be closed in the right half-plane.
Calculating the integrals in~(\ref{Eh-1}) by residues one finds
\begin{equation}\label{Eh-K}
\varphi_\xi(z)=
z^{-1}\sum_{j=2}^\infty \omega(j) A_\xi(j) \Psi_j^{\xi}(z)\,,
\end{equation}
where
\begin{eqnarray}
 A_\xi(j)&=&i\varkappa (-1)^{j-1}\,
\int_{-1}^{1}dx\, P_{j-1}\left(\frac{x}{|\xi|}\right)\,\partial_x\varphi_\xi(x)\nonumber\\[2mm]
&=&i\varkappa (-1)^{j}\,
\int_{-1}^{1}\frac{dx}{|\xi|}\, C^{3/2}_{j-2}\left(\frac{x}{|\xi|}\right)\varphi_\xi(x)\,.
\end{eqnarray}
The representation (\ref{Eh-K}) coincides
with~(\ref{Kiv}), which was obtained in
Ref.~\cite{KM-1}. Next, the sum in (\ref{Eh-1}) can be rewritten as an
integral over $j$
resulting in a representation similar to that derived in Ref.~\cite{BB}
\begin{equation}\label{Eh-B}
\varphi_\xi(z)=-\frac{i}{2z}\int_{C_a} \frac{dj}{\sin\pi j}
\omega(j) A^\pm_\xi(j) \Psi_j^{\xi}(z_\pm)\,,
\end{equation}
where
\begin{equation}\label{lmk1}
A_\xi^\pm(j)=i\varkappa\int_{-|\xi|}^1 dx \,P_{j-1}\left(\frac{x}{|\xi|}\right)\,
[\partial_x\varphi_\xi^\pm](\pm x)\,.
\end{equation}
The integration contour $C_a$ is shown in Fig.\ref{cont}. Deriving Eq.~(\ref{Eh-B})
we represented the function
$\varphi_\xi(x)$ as sum of two functions $\varphi_\xi^+(x)$ and $\varphi_\xi^-(x)$,
$\varphi_\xi(x)=\varphi_\xi^+(x)+\varphi_\xi^-(x)$, such that $\varphi_\xi^+(x)=0$ at
$x<-|\xi|$, and $\varphi_\xi^-(x)=0$ at
$x>|\xi|$.
We also notice that the contour
can not be deformed to be parallel to the imaginary axis, since in this case the integral
over $j$ starts to diverge.
All three expansions (\ref{Eh-1}),~(\ref{Eh-K}),~(\ref{Eh-B}) are equivalent, the difference
appears when one wants to make a Fourier transform. One can
interchange the Fourier transformation
with integration (summation) over $j$ only in the representation~(\ref{Eh-1}).
\vskip 0.5cm

Let us now show  that the expansion (\ref{Eh-1}) runs over the eigenfunctions of  the
Hamiltonian. First of all, we note that the Hamiltonian~(\ref{Ha})
is not a hermitian operator with
respect to the scalar product~(\ref{sc}). Then the commutativity,
$[\mathbb{H},\mathbf{J}^2]=0$, does not imply that the operators share the same
eigenfunctions. Indeed, one can easily check that the eigenfunctions of
the continuous spectrum~(\ref{Pcj})
of the Casimir operator are not the eigenfunctions of the Hamiltonian~(\ref{Ha}).

Nevertheless, one can find functions which diagonalize the Hamiltonian. (Of course, they are
not mutually orthogonal and do not form a basis of the Hilbert space.)
To solve the equation
$$
\mathbf{J}^2\varphi_j(z_1,z_2)\equiv-\partial_1\partial_2z_{12}^2\varphi_j(z_1,z_2)=
j(j-1)\varphi_j(z_1,z_2)\,
$$
we use the ansatz $\varphi_j(z_1,z_2)=e^{-i\xi(z_1+z_2)}\Phi^\xi_j(z_{12})$,
where $j$ is
an arbitrary
complex number.
This results in a  second order differential equation
for the function $\Phi^\xi_j(z)$. It has
two independent solutions
$\Phi^\xi_j(z)=z^{-1}\Psi^\xi_j(z)
$ and  $\Phi^\xi_{1-j}(z)=z^{-1}\Psi^\xi_{1-j}(z) $.
Due to commutativity of the integral operator
$\mathbb{H}$ and the differential operator $\mathbf{J}^2$
one concludes that
\be\label{HH}
[\mathbb{H}\,\varphi_j](z_1,z_2)=A(j)\,\varphi_j(z_1,z_2)+B(j)\,\varphi_{1-j}(z_1,z_2)\,,
\ee
whenever the action of the Hamiltonian $\mathbb{H}$ on the
function $\varphi_j(z_1,z_2)$ is well defined.
Substituting $\varphi_j(z_1,z_2)$ in the form $e^{-i\xi(z_1+z_2)}\Phi^\xi_j(z_{12})$
into  (\ref{Hq}) one finds that the integrals converge
when $\mathrm{Re}\ j>1$.
Next, let us notice that when
the variables $\alpha,\beta$  run over  the integration region,
the argument of the function $\Phi^\xi_j$ varies from $0$ to $z$.
Thus to fix  the coefficients $A(j)$ and $B(j)$ it is sufficient to study
the  asymptotics $z\to 0$ of the r.h.s and l.h.s. of (\ref{HH}). In this case one can
substitute
$\Phi^\xi_j(z)$ by its leading term
$\sim z^{j-2}$ \  and gets
$
B(j)=0
$
and $A(j)\equiv E(j)$ with
\begin{equation}\label{Ej}
E(j)=2\left[\psi(j)-\psi(2)\right]-\frac{1}{j(j-1)}+\frac12\,.
\end{equation}
Furthermore, we have to  fix the ambiguity in the definition of the function
$\Phi^\xi_j(z_{12})$  related to the existence of a branching
point at $z_{12}=0$.
To this end we note that the Hamiltonian~(\ref{HH}) transforms the functions
with  support in the region $z_{12}>0$ ($z_{12}<0$) into functions
with the same support property. Thus, we conclude, that  both functions,
$\theta(\pm z_{12})e^{-i\xi(z_1+z_2)}\Phi^\xi_j(\pm z_{12})$,
(or their linear combinations such as,
$e^{-i\xi(z_1+z_2)}\Phi^\xi_j(\pm z_{12}+i0)$)
are  eigenfunctions
of the Hamiltonian corresponding to the same eigenvalues, $E(j)$.
We notice also that since the integral
$\int dz z^2 |\Psi^\xi_j(\pm z)|^2$ is finite for \mbox{$\Re j>1/2$},
the eigenvalue $E(j)$ belongs to the discrete spectrum of the operator ${\mathbb H}$.

Thus we have shown that for any complex $j$ such that \mbox{$\Re j>1$} the function
\begin{equation}\label{eiHam}
\varphi_j^\pm(z_1,z_2)=e^{-i\xi(z_1+z_2)} z_{12}^{-1}\Psi^\xi_j(\pm z_{12}+i0)
\end{equation}
is an  eigenfunction of the discrete spectrum of
the Hamiltonian $\mathbb{H}$ with the eigenvalue given by
(\ref{Ej}).

\subsection{Solution of the evolution equation}

Since  we have shown that the expansion in the Eq.~(\ref{Eh-1}) contains eigenfunctions
of the Hamiltonian  one can easily write down the solution to
the evolution equation~(\ref{RGa})
in the LO approximation
\begin{align}\label{Evc}
\varphi_\xi^{\mu'}(z)=&\,
z^{-1}\left\{\sum_{j=2}^\infty\, \omega(j) \, a_\xi^{\mu}(j)\, L^{-\gamma(j)}
\Psi_j^{\xi}(z)\right.\nonumber\\[2mm]
&\left.-\sum_{a=\pm}
\frac{i}{2}\int_{\frac32-i\infty}^{\frac32+i\infty} \frac{dj}{\sin\pi j}\,
\omega(j)\, a^{\mu,a}_{\xi}(j)\,L^{-\gamma(j)}\, \Psi_j^{\xi}(z_a)\right\}\,,
\end{align}
where $\omega(j)=2j-1$,
\begin{eqnarray}
L=\frac{\alpha(\mu)}{\alpha(\mu')}& \text{   and  }
&\gamma(j)=2\frac{C_F}{b_0}E(j).
\end{eqnarray}
After Fourier
transform  Eq.~(\ref{Evc}) can be cast into the form
\begin{align}\label{Evp}
\varphi_\xi^{\mu'}(x)=
\sum_{j=2}^\infty\,  c_\xi^{\mu}(j)\, L^{-\gamma(j)}\,
p_j^{(1)}\left(\frac{x}{|\xi|}\right)~+~
\sum_{a=\pm} a
\int_C \frac{dj}{\pi i}\,
 c^{\mu,a}_{\xi}(j)L^{-\gamma(j)}q_j^{(1)}\left(\frac{ax}{|\xi|}\right),
\end{align}
where the integration follows the line $\Re j= k$, $1<k<2$.
The expansion coefficients are given by the following expressions
\begin{subequations}\label{coeff-q}
\begin{align}
\label{cd}
c_\xi^{\mu}(j)&=
v_j\int_{-|\xi|}^{|\xi|} dx P_{j-1}\left(\frac{x}{|\xi|}\right)
\,\partial_x \varphi_\xi^{\mu}(x)\,,\\[2mm]
\label{cc}
c_\xi^{\mu,\pm}(j)&=
v_j
\int_{|\xi|}^1 dx P_{j-1}\left(\frac{x}{|\xi|}\right)
\,[\partial_x \varphi_\xi^{\mu}](\pm x)\,,
\end{align}
\end{subequations}
where $P_{j-1}(x)$ are the Legendre functions of the first kind and $v_j=(2j-1)/2$.
Notice, that the coefficients $c_\xi^{\mu,\pm}(j)$ are antisymmetric under the
interchange
$j\to 1-j$, $c_\xi^{\mu,\pm}(j)=-c_\xi^{\mu,\pm}(1-j)$.
We also want to stress here that the coefficients $c_\xi^{\mu,\pm}(j)$ are  entire
functions of $j$ in the whole complex plane.

The functions $p_j^{(1)}(x)$, $q_j^{(1)}(x)$ are expressed in terms of the Legendre
functions of the first and second kind~\cite{GR}.
We give here the expressions for the functions $p_j^{(m)}(x)$, $q_j^{(m)}(x)$
for  general integer $m$ since they appear in the solution of  the
evolution equation in
the singlet sector. These functions (up to some normalization factors)
are the Fourier transform of the functions $z^{-m} \Psi_j^{\xi}(z)$  ($p^{(m)}_j(x)$) and
$z^{-m} \Psi_j^{\xi}(z+i0)$ ($q^{(m)}_j(x)$). The function $ p^{(m)}_j(x)$
is defined as follows
\begin{align}\label{pj}
p_j^{(m)}(x)&&
=(-1)^m\theta(1-x^2)(1-x^2)^{m/2} P_{j-1}^{-m}(x)
=
r_j^m\theta(1-x^2)(1-x^2)^m C_{j-m-1}^{m+1/2}(x)\,,
\end{align}
where
$$
r_j^m=(-1)^m2^{-m}\frac{\Gamma(2m+1)}{\Gamma(m+1)}\frac{\Gamma(j-m)}{\Gamma(j+m)}.
$$
The function $q_j^{(m)}(x)=0$ for $x<-1$, while for $x>-1$ it is given by
\begin{equation}\label{qj}
q_j^{(m)}(x)=(x^2-1)^{m/2}Q_{j-1}^{-m}(x),\ \ \ \ x>1
\end{equation}
\begin{equation}\label{qjc}
q_j^{(m)}(x)=\frac{\pi}{2\sin\pi j} (1-x^2)^{m/2} P_{j-1}^{-m}(-x)\,, \ \ |x|<1\,.
\end{equation}
The function $q_j^{(1)}(x)$ is continuous at the point $x=1$,
\begin{equation}	
q_j^{(1)}(x)\bigl|_{x=1}=-\frac1{j(j-1)}\,,
\end{equation}	
but its first
derivative has a logarithmic singularity at this point
\begin{equation}
\frac{d}{dx} q_j^{(1)}(1+x)\sim -\frac12 \log x+\ldots\,.
\end{equation}

The formula~(\ref{Evp}) represents the solution to the evolution equation
for the pion isovector quark GPD, $H(x,\xi)\equiv \varphi_\xi(x)$.
It can be used both for numerical and analytical study of the evolution.
\vskip 0.5cm

Let us discuss the solution~(\ref{Evp}) in more details.
First of all we notice that the integrals over $j$ in (\ref{Evp}) vanish whenever
$|x|>1$.
Second,
at the input scale $\mu'=\mu$ \mbox{($L=1$)} the Eq.~(\ref{Evp}) can
be represented as an expansion
in eigenfunctions of the Casimir operator. To this end we shift
the integration contour to the line $\Re j=1/2$.
Since the function $q_j^{(1)}(x)$ for $|x|<1$ has a pole
at $j=1$ one should calculate the residue at the point $j=1$.
Taking into account that
\begin{equation}\label{boundary}
c^{\mu,\pm}_\xi(j)|_{j=1}=\mp\frac12\varphi_\xi^\mu(\pm|\xi|)
\end{equation}
and
$
(1-x^2)^{1/2}P_0^{-1}(-x)=(1+x)
$
one finds that the contribution from residues is
\begin{align}\label{p0}
\Delta^\mu(x,\xi)=\frac12\theta(|\xi|-|x|)
\left[\left(1+\frac{x}{|\xi|}\right)\varphi_\xi^\mu(|\xi|)+
\left(1-\frac{x}{|\xi|}\right)\varphi_\xi^\mu(-|\xi|)\right]\,.
\end{align}
Using the antisymmetry of the expansion coefficients,
$c^{\mu,a}_{\xi}(j)$, under $j\to 1-j$ and the relation (\ref{PQ}) between the Legendre
functions
one can rewrite  Eq.~(\ref{Evp})  in the form
\begin{align}\label{E0}
\varphi_\xi^\mu(x)=\Delta^\mu(x,\xi)+\sum_{j=2}^\infty  c_\xi^{\mu}(j)
p_j^{(1)}\left(\frac{x}{|\xi|}\right)+
\sum_{a=\pm}\frac{a}{2 i}\int_{1/2-i\infty}^{1/2+i\infty} dj\,\cot\pi j\,
 c^{\mu,a}_{\xi}(j) {\mathcal P}_j^{(1)}\left(\frac{ax}{|\xi|}\right),
\end{align}
where
\begin{equation}\label{PP-1}
{\mathcal P}_j^{(m)}(x)=\theta(x-1)\,(x^2-1)^{m/2} P_{j-1}^{-m}(x)\,.
\end{equation}
Thus, one sees that the contribution of the integrals in Eq.~(\ref{Evp}) at $L=1$
in the region $|x|<|\xi|$ is given by the term $\Delta_\xi^\mu(x)$, which
is determined completely by the values of the function at the points $x=\pm \xi$.
Obviously,  the first two terms in (\ref{E0}) have  support in the ERBL
region, while the integrals live in the DGLAP region. The term
$\Delta_\xi^\mu(x)$
can be restored from the knowledge of the function in the DGLAP region.

At the scale $\mu'$ the function $\varphi_\xi(x)$ is given by the expression~(\ref{Evp}).
To rewrite it in the form~(\ref{E0}) one has to find the expansion coefficients
$c_\xi^\mu(j)$ and $c^{\mu,\pm}_\xi(j)$ at the scale $\mu'$.
To this end one should insert the expansion (\ref{Evp})
into (\ref{coeff-q}) and evaluate the corresponding integrals.
This results in the following expression for the expansion coefficients
at scale~$\mu'$
\begin{subequations}\label{cnew}
\begin{align}
\label{cc-new}
c_\xi^{\mu',\pm}(j)=&
\frac{v_j}{\pi i}\int_{C_{II}} dj'
\frac{L^{-\gamma(j')}c^{\mu,\pm}(j')}{(j'-j)(j'+j-1)}\,,\\[2mm]
\label{cd-new}
c_\xi^{\mu'}(j)=&c_\xi^\mu(j)L^{-\gamma(j)}-
\frac{v_j}{\pi i}\int_{C_I} dj'L^{-\gamma(j')}
\frac{c^{\mu,+}(j')+(-1)^{j-1}c^{\mu,-}(j')}{(j'-j)(j'+j-1)}\,,
\end{align}
\end{subequations}
where the integration contour in both cases follows the line parallel to the
imaginary axis such that $1<\Re j'<j$ (contour $C_I$),
and  $\max\{1,\Re j, 1-\Re j\}<\Re j'$ (contour $C_{II}$).
We notice that the integration contours can not be closed in the right half-plane
because the integrals over the large semicircle do not vanish.
Next, let us check that at
$\mu'=\mu$ ($L=1$) the r.h.s reproduces the l.h.s.
To this end let us shift the integration contour
to the line $\Re j'=1/2$. Then the r.h.s of  Eq.~(\ref{cc-new}) will be given
by the sum of  the residue at the point $j'=j$ and the integral over the line $\Re j'=1/2$.
The integral vanishes due to antisymmetry of the integrand under  $j\leftrightarrow 1-j$
while the residue gives the coefficient $c_\xi^{\mu,\pm}(j)$.

One sees that the coefficients $c_\xi^\mu(j)$, $c^{\mu,\pm}_\xi(j)$ mix under
evolution~\footnote{The coefficients $c^{\mu,\pm}_\xi(j)$ are the
matrix elements of the so-called
string operators~\cite{BB}.}. Therefore, their scale dependence
is not independent.  Nevertheless, it is straightforward to check that the
transformation $c^\mu\to c^{\mu'}=f(\mu',c^\mu)$ generated by Eqs.(\ref{cnew}) possesses the
necessary group property, $f(\mu'',c^\mu)=f(\mu'',f(\mu',c^\mu))$.

The coefficients $c_\xi^{\mu'}(j)$ get  contributions from
$c_\xi^{\mu,\pm}(j)$. This effect can be interpreted as  migration
of partons from the DGLAP region to the ERBL
region~\cite{Radyushkin96,RB}.

Further, let us note also that it follows from the Eq.(\ref{cnew})
that for integer $j\geq 2$
\begin{multline}\label{CG}
\left[c_\xi^{\mu'}(j)+c_\xi^{\mu',+}(j)+(-1)^{j-1}c_\xi^{\mu',-}(j)\right]=
\left[c_\xi^{\mu}(j)+c_\xi^{\mu,+}(j)+(-1)^{j-1}c_\xi^{\mu,-}(j)\right]
L^{-\gamma(j)}\,.
\end{multline}
It is not surprising that the combination in the brackets is  nothing
but
the Gegenbauer
moment of the GPD
\begin{equation}\label{CNN}
C_{j-2}^\mu(\xi)=-v_j\int_{-1}^1\frac{dx}{|\xi|}
 \,C_{j-2}^{3/2}\left(\frac{x}{|\xi|}\right)\,\varphi_\xi^\mu(x)\,.
\end{equation}
For the physical GPD, $H(x,\xi)$, moments $\xi^N C_{N}^\mu(\xi)$ have to be polynomials
in $\xi$ of degree $N$. This is the so-called polynomiality conditions~\cite{Ji96}.
It means that the knowledge of the GPD in the DGLAP region allows
to restore the GPD in the central region, up to  terms which satisfy the polynomiality
conditions.

Next, all singularities of the integrand in the integral in  Eq.~(\ref{cc-new})
lie to the left of the integration line. This allows to shift it arbitrarily far
to the right. Since the anomalous dimension $\gamma(j)$ behaves as $\log j$ for large $j$
we conclude that the coefficients $c^{\mu,\pm}_\xi(j)$ vanish faster than any power of $1/L$
for $L\to \infty$. As far as the function $\varphi_\xi(x)$ 
in the DGLAP region $|x|\geq|\xi|$ is expressed solely
in terms of the coefficients $c^{\mu,\pm}_\xi(j)$ this implies also that the function in this
region vanishes faster any power of $1/L$
for $L\to \infty$.
Thus the asymptotic expansion  for the function
$\varphi_\xi^{\mu'}(x)$ for $\mu'\to\infty$ (large $L$) has the form
\begin{equation}\label{asf}
\left[\varphi_\xi^{\mu'}(x)-\sum_{j=2}^n C_{j-2}^\mu\,L^{-\gamma(j)}\,
p^{(1)}_j\left(\frac{x}{|\xi|}\right)\right]\sim {\cal O}\left(L^{-\gamma(n+1)}\right)\,.
\end{equation}
Since the anomalous dimension $\gamma(j)$ vanishes for $j=2$ 
the function $\varphi_\xi^{\mu'}(x)$ tends
to its asymptotic form, $p^{(1)}_2\left({x}/{|\xi|}\right)\sim
(1-x^2/\xi^2)$, (see Refs.~\cite{RB,BGMS}), 
for $\mu'\to\infty$.
We stress that the expansion~(\ref{asf}) is an asymptotic expansion and that the
series in (\ref{asf}) diverges for any $x$.

\begin{figure}[t]
\psfrag{1}[cc][cc][0.8]{$(1)$}
\psfrag{2}[cc][cc][0.8]{$(3)$}
\psfrag{3}[cc][cc][0.8]{$(2)$}
\psfrag{4}[cc][cc][0.8]{$(4)$}
\psfrag{x}[cc][cc][1.2]{$x$}
\psfrag{f}[lc][rc][1.2]{$\varphi_\xi(x)$}
\centerline{\epsfxsize8.5cm\epsfbox{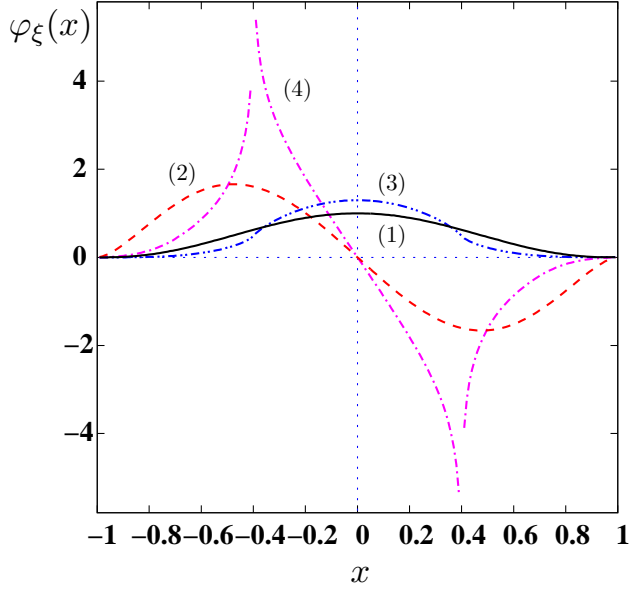}}
\caption[]{Example for the evolution of the function
$\varphi_{\xi=0.4}(x)=(1-x^2)^{2.7}$ (curve $(1)$)
and its derivative (curve $(2)$)
to the scale $L=1.5$,
curves $(3)$ and $(4)$, respectively.}
\label{der-ex}
\end{figure}
Let us analyze now the behavior of the function $\varphi_\xi^{\mu'}(x)$
at the point $x=\pm \xi$.  As follows from the properties of the functions $q_j^{(1)}(x)$
the GPD $\varphi_\xi^{\mu'}(x)$ is continuous at the points $x=\pm \xi$ and has
the following values (see Eqs.~(\ref{boundary}) and (\ref{cc-new}))
\begin{equation}
\varphi_\xi^{\mu'}(\pm |\xi|)=  \mp\frac{1}{\pi i}\int_{\kappa-i\infty}^{\kappa+i\infty} dj'
\frac{c^{\mu,\pm}(j')}{j'(j'-1)}\,L^{-\gamma(j')}\,,
\end{equation}
where the $\kappa>1$.
Taking into account the Eqs.~(\ref{pqap}) and (\ref{Qz}) one finds that the
 first derivative of the function $\varphi_\xi^{\mu'}(x)$ has at least a
  logarithmic singularity at
$x=\pm \xi$ even if  the function $\varphi_\xi(x)$ is a smooth function at the input scale $\mu$
(see Fig.~\ref{der-ex})
\begin{equation}
\frac{d}{dx}\varphi_\xi^{\mu'}(x)\underset{x\to \pm \xi}{\sim}
\alpha_\pm |\xi|^{-1}\log\left|{x}/{|\xi|}\mp 1\right|\,.
\end{equation}
Here
\begin{equation}\label{alpha}
\alpha_\pm=-\frac{1}{2\pi i}  \int_{\kappa-i\infty}^{\kappa+i\infty} dj'
{c^{\mu,\pm}(j')}\,L^{-\gamma(j')}\,.
\end{equation}
If the integral in~(\ref{alpha}) diverges it means that the derivative of
$\varphi_\xi^{\mu'}(x)$ has a more singular behaviour at $x=\pm
\xi$. It is clear that the
convergence properties of the integral in (\ref{alpha}) is determined by the behavior of the
coefficients $ c^{\mu,\pm}(j) $ for $j\to \pm i\infty$. The latter depends on the
smoothness of the function $\varphi_\xi(x)$ at the scale $\mu$, i.e. on the behavior of
the function at the points $x=\pm\xi$ and $\xi=\pm 1$.
\vskip 0.5cm

The amplitude of a scattering process (DVCS, light meson production, etc.) is given
by the convolution of a GPD with a hard scattering amplitude. The typical integral
occurring in such a convolution at lowest order in $\alpha_s$ is
\begin{equation}
{\mathcal J}(\mu',\xi)=\int_{-\infty}^{\infty} dx \frac{\varphi_\xi^{\mu'}(x)}{\xi- x-i0}\,,
\end{equation}
where due to the support properties of the GPD
the integration is restricted to the interval $[-1,1]$.
We assume here  that $\xi>0$.
Substituting $\varphi_\xi(x)$ by  the expansion (\ref{Evp}) and changing the order of
summation (integration) over $j$ and $x$ one  derives with the help of the Eqs.~(\ref{pq-int})
\begin{align}
{\mathcal J}(\mu',\xi)=-2\sum_{j=2} \frac{c_\xi^{\mu}(j)}{j(j-1)}\,
L^{-\gamma(j)}+
\frac{1}{\pi i}\int_C
\frac{dj}{\sin\pi j}\,
\frac{\left[ e^{-i\pi j}
c^{\mu,+}_{\xi}(j)-c^{\mu,-}_\xi(j)\right]}{j(j-1)}L^{-\gamma(j)}\,,
\end{align}
where the integration follows the  line $\Re j=\kappa$, $1<\kappa<2$.
\vskip 0.5cm

Let us compare the approach presented here with the one developed
in~\cite{MS}.
One can see that the only difference between Eq.~(81) in
Ref.~\cite{MS} and our representation of the GPD,  i.e.
Eq.~(\ref{Evp}) at $L=1$, 
(or analoguosly Eq.~(4) in Ref.~\cite{MKS}) is, that the sum over
discrete $j$ in Eq.~(\ref{Evp}) is rewritten
as an integral over $j$. To this end one has to construct the
analytical continuation
of the coefficients $c_\xi(j)$ with suitable analytic properties. This
has been done
in~\cite{MS}. However,  Eq.~(69) in Ref.~\cite{MS}, which provides
such an analytical continuation, 
involves the function $\varphi_\xi(x)$ (or $w(x,\eta)$ in the
notations of Ref.~\cite{MS}) outside the region where it is
originally defined. 
Thus, such reformulation is possible only under certain additional
assumptions on the analytic structure of GPDs.

\subsection{Small $\xi$ asymptotics}
In this subsection
we want to  discuss the structure of the solution (\ref{Evp}) in the limit
$\xi\to0$. (It is clear that when $\xi\to1$ one recovers the
conventional ERBL
evolution~\cite{LB,ER}.)
Obviously, the form of the function in the small $\xi$
region can not differ strongly from the input function at $L\sim 1$. However,
one may hope that the form of the evolved
function for large scales, $L\gg 1$, depends weakly on the initial profile.

The sum in  Eq.~(\ref{Evp}) contributes only to the central region. To estimate
this sum one has to have some information on the expansion
coefficients $c_\xi(j)$.
If the coefficients $c_\xi(j)$ are not singular as $\xi\to 0$, then at large $L$ one can
omit all the terms in sum except the first one. This type of behavior
holds for some models
of GPDs~\cite{Rad98,Musatov} but the opposite situation is not
excluded.
Thus, we restrict ourself
to the analysis of the small $\xi$ behavior of the integrals in Eq.~(\ref{Evp}).

First of all we shall show, that in the limit $\xi\to 0$ and $x$ fixed, Eq.~(\ref{Evp}) takes
the standard DGLAP form. Till the end of this section we shall assume that $\xi>0$.
Let us consider  the integral in (\ref{Evp}) corresponding to  the
term ``$a=+$''.
For $\xi\to 0$ and $x$ fixed,
the argument of the function  $q_j^{(1)}$ goes to infinity and so we
can replace it by its
asymptotic value
\begin{align}
q_j^{(1)}(x/\xi)\underset{\xi\to 0}{=}&-x^{1-j}\frac{\sqrt{\pi}\Gamma(j-1)}{2\Gamma(j+1/2)}
\left(\frac{\xi}{2}\right)^{j-1}\left(1+{\mathcal O}({\xi^{2}})\right)\,.
\end{align}
The contour of the integration over $j$  follows the line $\Re j=\kappa$,
 $\kappa>1$. The function $q_j^{(1)}$ vanishes as $\xi^{\kappa-1}$ for $\xi\to 0$
and the parameter $\kappa$ can be chosen sufficiently large.  In its turn the coefficient
$c^+_\xi(j)$ becomes singular at $\xi\to 0$.  Indeed, using  Eq.~(\ref{P1i}) one can
derive the following representation
\begin{eqnarray}\label{Mc}
c^+_\xi(j)=-\frac{v_j}{4\pi^{3/2} i}
\int_C ds \,\varphi_\xi(s) 
\left(\frac{\xi}{2}\right)^{-s}\frac{\Gamma\left(\frac{s+j}{2}\right)
\Gamma\left(\frac{s-j+1}{2}\right)}{\Gamma(s)}\,,
\end{eqnarray}
where the  integration goes from $-i\infty$ to $+i\infty$  staying to the right
of all singularities of the integrand and
\begin{equation}\label{Mln}
\varphi_\xi(s)=\int_0^1 dx\, x^{s-1}\, \varphi_\xi(x)\,.
\end{equation}
If $\Re s $ is sufficiently large  then $\varphi_\xi(s)\to \varphi_{\xi=0}(s)$ for $\xi\to 0$
even if the function $\varphi_\xi(x)$ is singular at \mbox{$x=\xi=0$.}
The leading $\xi\to 0$ asymptotics of  $c^+_\xi(j)$ is determined by the rightmost
singularity of the integrand in (\ref{Mc}) which is situated at $s=j-1$. Thus one gets
\begin{align}\label{cpa}
c^+_\xi(j)=-\left(\frac{\xi}{2}\right)^{1-j}\frac{\Gamma(j+1/2)}{\sqrt{\pi}\Gamma(j-1)}
\varphi_\xi(j-1)\left(1+{\mathcal O}(\xi^2)\right)\,.
\end{align}
Therefore, for $\xi=0$ the solution (\ref{Evp}) ($x>0$) takes the DGLAP form
\begin{equation}\label{AP}
\varphi_{\xi=0}^{\mu'}(x)=\frac1{2\pi i}\int_C dj\, x^{-j}\,\varphi_{\xi=0}(j)\,
L^{-\gamma(j+1)}\,.
\end{equation}

The integrals in (\ref{Evp}) depend on the three parameters $x,\xi$ and $L$.
Eq.~(\ref{AP}) was obtained in the limit $\xi\to 0$, $x$ and $L$ fixed. Let us now
keep the ratio $x/\xi=r>1$ fixed (i.e. we shall consider the  DGLAP region)
and evaluate the integral
\begin{equation}\label{Int}
{\mathcal I}(\xi,L,r)=\frac{1}{\pi i}
\int_C {dj}\,
 c^{\mu,+}_{\xi}(j)L^{-\gamma(j)}q_j^{(1)}(r)
\end{equation}
for large  $1/\xi$ and $L$. The answer  depends on the relative size of $L$ and
$\xi$. Lets start with the situation when $\log L\gg \log 1/\xi$. In this case taking into
account  Eq.~(\ref{cpa}) one can evaluate the integral by the
saddle point method.
The position of the saddle point is determined from the equation
\begin{equation}\label{saddle}
1-\gamma'(j_{*}) \sigma=0\,, \ \ \ \ \sigma=\log L/\log\, (2/\xi)\,.
\end{equation}
For large $\sigma$, (which of course corresponds to an unphysically high scale $\mu'$)
$j_*=4C_F/b_0\,\sigma$
and
the integral~(\ref{Int}) scales with $L$ as
$$
{\mathcal I}\sim L^{-4C_F/b_0 (\log j_*-1)} q_{j_*}^{(1)}(r).
$$
Note that we restricted our consideration to  the DGLAP region, $x/\xi>1$, only because
the integrand in this region has no singularities in the half-plane $\Re j>1$, so that that
 we can
freely deform the  integration contour. In the ERBL region the integrand has
poles, (see Eq.~(\ref{qjc})), so when deforming contour one has to take into account the
 corresponding
residues. These residues together with the terms in the sum in (\ref{Evp})
result in the asymptotic expansion (\ref{asf}). Consequently,
the number of  terms in the asymptotic
expansion (\ref{asf}) is controlled by the position of the saddle
point $j_*\sim \log L/\log \,(2/\xi)$.
This means that the ``time'' $\tau=\log L$ which is necessary
for the function to reach the asymptotic
regime is proportional to $\log 2/\xi$.

Now let us decrease $L$ or $\xi$. Then $\sigma$ and $j_*$ decrease as well.
When $\sigma\ll 1$ (i.e. when the function is far from its asymptotic form) one finds
\begin{equation}
j_*\simeq 1+\sqrt{\frac{2C_F}{b_0}\sigma}\,
\end{equation}
and recovers the well known ``double scaling'' behaviour~\cite{RU,SF}
\begin{align}
{\mathcal I}
\sim
\frac{\log^{1/4} L }{\log^{3/4}(2/\xi)}\, e^{\sqrt{8C_F/b_0 \log L\log 2/\xi}}\,
q_{j_*}^{(1)}(r).
\end{align}

Let us now consider the situation when the function
$\varphi_{\xi=0}(x)$ is enhanced at small $x$. Recent fits of parton
densities~(see Ref.~\cite{CTEQ}) shows that in the $x$ region, $10^{-4}\leq x\leq  10^{-2}$,
they display a power-like behavior $\varphi_{\xi=0}(x)\sim x^{-\alpha}$.
In this case the Mellin transform of the function $\varphi_{\xi=0}(x)$ has a pole
at $s=\alpha$. (In the case $\alpha>0$ the first pole of $\varphi_{\xi=0}(j-1)$ lays
to the right of the  first pole of the anomalous dimension $\gamma(j)$.)
At the same time  the function $\varphi_{\xi}(x)$ is a regular function
of $x$ for $\xi\neq 0$, so that the first pole of its Mellin transform lies at $s=0$.
It can be seen that near the point $s=\alpha$ the Mellin transform $\varphi_\xi(s)$
takes the following form $\varphi_\xi(s)\sim (1-\xi^{s-\alpha} f(\xi,s))/(s-\alpha)$,
where $f(\xi,\alpha)=1$. One sees that the numerator contains two terms with  different
$\xi$ dependence. Thus the integrand in (\ref{Int}) near the point $j=1+\alpha$ can be
split up into two terms, which are proportional to  $\xi^{1-j}$ and 
$\xi^{-\alpha}$, respectively.
Each of these terms has  poles, which, of course, cancel in the sum.
Calculating the asymptotics of the first term ($\xi^{1-j}$) one finds that in the case
 $j_*<1+\alpha$ it is necessary to take into account the contribution due the pole
at $j=1+\alpha$ which is
\begin{equation}\label{1-term}
{\mathcal I}^{pole}\sim\xi^{-\alpha}\, q_{1+\alpha}^{(1)}\left(\frac{x}{|\xi|}\right)\,
L^{-\gamma(1+\alpha)}\,.
\end{equation}
For $0<\alpha<1$ the anomalous dimension $\gamma(1+\alpha)$ is
negative so that the evolution results in an enhancement of this contributions.
The second term has the same  $\xi$
dependence as the pole contribution, $\xi^{-\alpha}$.
However, this term vanishes  faster with $L$ than the first one. So one can expect that
the pole contribution (\ref{1-term}) will dominate in the limit
$L\to\infty$, $\log L/\log 1/\xi$ fixed.
 Our result~(\ref{1-term}) agrees with the result
of Shuvaev et al.~\cite{SGMR}.
(One can easily check that the integral~(22) in Ref.~\cite{SGMR} gives rise
to the function $ q_{2+\lambda_i}^{(p+1)}\left({x}/{\xi}\right)$.)

\setcounter{equation}{0}
\section{Singlet case}\label{singlet}
In the singlet case one is dealing with the evolution of the quark,
$\mathcal{F}^{q,+}(x,\xi)$,
and gluon, $\mathcal{F}^{g}(x,\xi)$, GPDs  which are related to 
matrix elements of the $C-$even
quark and gluon operators, (see Eqs.~(\ref{Fqg}), (\ref{Fq+})). We shall proceed along the
same lines as in the case of the isovector quark GPD.  Similarly, we start with the
formulation of the problem in coordinate space. The only difference with respect to
the previous
case is that we have now two functions $f^{q}(z_1,z_2)$, $f^{g}(z_1,z_2)$ (see Eqs.~(\ref{fGF}))
instead of one ($\varphi(z_1,z_2)$).  For the sake of brevity we
shall denote by
$f(z_1,z_2)$ the two component function
$f(z_1,z_2)=(f^{q}(z_1,z_2),f^{g}(z_1,z_2))\equiv
(f^{(1)}(z_1,z_2),f^{(2)}(z_1,z_2))$.
Next, let us introduce a function $\psi$,
$
\psi=(\psi^q,\psi^g)\equiv (\psi^{(1)},\psi^{(2)})\,,
$
as follows
\begin{subequations}
\begin{eqnarray}\label{map-2}
\psi^{(k)}(z_1,z_2)&=&(z_1-z_2)^k f^{(k)}(z_1,z_2)\,,
\end{eqnarray}
\end{subequations}
where $k=1,2$. 
The functions $\psi^{q(g)}$ transform under $SL(2,R)$ transformations according to
 Eq.~(\ref{t-q}) with $s=1/2$ while the functions $f^{q(g)}$ obey the same transformation
with $s=1$ and $s=3/2$, respectively.
We define the scalar product on the space of functions $f$ as follows
\begin{equation}
\vev{f_1|f_2}\equiv \vev{\psi_1|\psi_2}=
\vev{\psi^{g}_1|\psi^{g}_2}+\vev{\psi^{q}_1|\psi^{q}_2}\,,
\end{equation}
where the scalar product $\vev{\psi_1^{\alpha}|\psi_2^{\alpha}}$ is the
standard scalar product on the space $\mathbf{L}^2(R\times R)$ , (see  Eq.~(\ref{l2r})).

Since the functions  $\psi^{q(g)}$ obey the  $SL(2,R)$
transformations~(\ref{t-q})
with conformal spin $s=1/2$ they can be expanded in eigenfunctions of the Casimir
operator according to the Eq.~(\ref{Exp12}). Again, we shall omit the Fourier integral
over $\xi$ and write down all expansions for the function $f_\xi(z)$ ($\psi_\xi(z)$)
\begin{equation}
f(z_1,z_2)=\int \frac{d\xi}{2\pi} e^{-i\xi(z_1+z_2)}\,f_\xi(z_1-z_2)\,.
\end{equation}
Obviously, an expansion for the function $\psi^k_\xi(x)$ has the form~(\ref{Eh}).
Then we  shift the integration contour from
the line $\Re j=1/2$ to $\Re j = 5/2$. Doing so one has to take
into account the residues
at the points $j=1,2$ which cancel identically with the first two
terms in the sum over $j$.
To see that the sum of the residues at $j=2$ for the function $\psi^q_\xi(z)$ vanishes
one should remember that this function is symmetric with respect to $z\to -z$.
Thus the expansion for the function $f_\xi(z)$ can be written in the following form
\begin{align}\label{ES-1}
f_\xi^{(k)}(z)=\sum_{j=3}^\infty\, \omega(j)\, a_\xi^{(k)}(j)\,\,
\Psi_j^{\xi,(k)}(z)~-
\frac{i}{2}\int_{\kappa-i\infty}^{\kappa+i\infty}
\frac{dj}{\sin\pi j}\,
 {\omega(j)}\,a^{(k),\pm}_\xi(j)\,\Psi_j^{\xi,(k)}(z_\pm) \,,
\end{align}
where $2<\kappa<3$, $k=1,2$, $z_\pm=\pm z+i0$ and
the coefficients $a_\xi^{(k)}(j),a^{(k),\pm}_\xi(j)$  are given by Eq.~(\ref{exc-p}) after
the substitution $\psi\to \psi^{(k)}$. The functions $\Psi_j^{\xi,(k)}(z_\pm)$ are
defined as follows
$$
\Psi_j^{\xi,(k)}(z_\pm)=z^{-k}\Psi_j^{\xi}(z_\pm)\,,\ \ \
$$
(and similar for $\Psi_j^{\xi,(k)}(z)$)
where $\Psi_j^{\xi}(z)$ is defined in Eq.~(\ref{psij}).
Now, using the same arguments as in  Sect.~\ref{s-q-gpd} one can show that
the Hamiltonians
describing the renormalization of the singlet operators~ Eqs.~(\ref{Hgg})--(\ref{H-21})
act on the functions
$\Phi^{(k)}_j(z_1,z_2)=e^{-i\xi(z_1+z_2)} \Psi^{\xi,(k)}_j(z_{12})$
as follows
\begin{equation}
{\mathcal H}^{ik}\,\Phi^{(k)}_j(z_1,z_2)=E^{ik}(j)\,
\Phi^{(i)}_j(z_1,z_2)\,.
\end{equation}
Note that in the formula above no summation over repeated indices
$i$, $k$ is implied.
The matrix of the "anomalous dimensions" $E^{ik}$ has the well known form
\begin{subequations}
\begin{align}
E^{qq}(j)\equiv
E^{11}(j)
=&C_F\left(2\left[\psi(j)-\psi(2)\right]-\frac{1}{j(j-1)}+\frac12\right)\,,\\[2mm]
E^{qg}(j)\equiv
E^{12}(j)
=&n_f\frac{j^2-j+2}{(j+1)j(j-1)(j-2)}\,,\\[2mm]
E^{gq}(j)\equiv
E^{21}(j)
=&C_F\frac{j^2-j+2}{j(j-1)}\,,\\[2mm]
E^{gg}(j)\equiv
E^{22}(j)
=&2N_c\left[\psi(j)-\psi(3)-\frac{1}{j(j+1)}
-\frac{1}{(j-1)(j-2)}\right]+\frac{7}{6}N_c+\frac13n_f\,.
\end{align}
\end{subequations}
The solution of the evolution equation (\ref{RGG}) for the function $f_\xi(z)$
can be written in the form
\begin{align}\label{sev-z}
f_\xi^{(k),\mu'}(z)&
=\sum_{j=3}^\infty\, \omega(j)\,
\left(L^{-\gamma(j)}a_\xi^\mu(j)\right)_k\,\Psi_j^{\xi,(k)}(z)
\nonumber\\[2mm]
&-\frac{i}{2}
\int_C
\frac{dj}{\sin\pi j}\,
 {\omega(j)}\,\left(L^{-\gamma(j)}a_\xi^{\mu,\pm}(j)\right)_k\,
 \Psi_j^{\xi,(k)}(z_\pm)\,.
\end{align}
Here, $\gamma(j)$ is the matrix of the anomalous dimensions, $\gamma(j)=2E(j)/b_0$,
$a^\mu_\xi(j)$ is a two-dimensional vector,\
 $a^\mu_\xi(j)=(a^{\mu,(1)}_\xi(j),a^{\mu,(2)}_\xi(j))$,\
and the integration contour is the same as in Eq.~(\ref{ES-1}).

By taking the Fourier transform  Eq.~(\ref{sev-z}) can be
brought into the form
\begin{align}\label{Ev-S}
\vec{\mathcal F}^{\mu' }(x,\xi)
=
\sum_{j=3}^\infty\, {\widehat p}_j\left(\frac{x}{|\xi|}\right)L^{-\gamma(j)} \,
{\vec{c}\,}_\xi^{\mu}(j)\,
+
\sum_{a=\pm}
\int_C \frac{dj}{\pi i}\, {{\widehat q}_j}^{\,\,a}\left(\frac{ax}{|\xi|}\right)
L^{-\gamma(j)}\,\vec{c\,}^{\mu,a}_{\xi}(j)\,.
\end{align}
The quark and gluon GPDs
  ${\mathcal F}_{1(2)}^\mu(x,\xi)$ are defined in  Eq.~(\ref{FF}).
The matrix functions $\widehat p_j(x)$ and  $\widehat q_j(x)$
are given by the following expressions
\begin{eqnarray}
\widehat p_j(x)&=&\left(\begin{array}{cc} p_j^{(1)}(x)&0\\ 0&\xi\, p_j^{(2)}(x)
\end{array}\right)\\[2mm]
{\widehat q_j}^{\,\,\pm}(x)&=&\left(\begin{array}{cc} \pm q_j^{(1)}(x)&0\\ 0&\xi\, q_j^{(2)}(x)
\end{array}\right)
\end{eqnarray}
and the functions $p_j^{(m)}, q^{(m)}_j$ are defined by   Eqs.~(\ref{pj}),~(\ref{qj}).
The expansion coefficients, $\vec{c}=(c^{(1)},c^{(2)})$ have the following form
\begin{subequations}\label{ccs}
\begin{align}
\label{ccsd}
c^{(k),\mu}_\xi(j)&=v_j\int_{-|\xi|}^{|\xi|} dx P_{j-1}\left(\frac{x}{|\xi|}\right)
\left[\frac{d^k}{dx^k}
{\mathcal F}_k^\mu\right](x,\xi)\,,\\[2mm]
\label{ccsc}
c^{(k),\mu,\pm}_\xi(j)&=v_j\int_{|\xi|}^{1} dx P_{j-1}\left(\frac{x}{|\xi|}\right)
\left[\frac{d^k}{dx^k}
{\mathcal F}_k^\mu\right](\pm x,\xi)\,.
\end{align}
\end{subequations}
Note, that the gluon GPD ${\mathcal F}_2(x,\xi)\equiv  {\mathcal F}_g(x,\xi)$ is a  symmetric
function of $x$, while the quark GPD
${\mathcal F}_1(x,\xi)\equiv  {\mathcal F}_q(x,\xi)-{\mathcal F}_q(-x,\xi)$
is an antisymmetric one. Therefore, all integer moments with even
$j$ (Eq.~(\ref{ccsd})) vanish  and $\vec{c\,}^{\mu,+}_\xi(j)=\vec{c\,}^{\mu,-}_\xi(j)$.
The matrix exponent $L^{-\gamma(j)}$ can be written in the form
\begin{equation}\label{Lexp}
L^{-\gamma(j)}=L^{-\gamma_+(j)} A_+(j)+L^{-\gamma_-(j)} A_-(j)\,,
\end{equation}
where
\begin{align}
\gamma_\pm(j)=&
\frac{\gamma_{11}(j)+\gamma_{22}(j)}{2}\mp\Delta(j)\,,
\\[2mm]
\Delta(j)=&\sqrt{\frac{(\gamma_{11}(j)-\gamma_{22}(j))^2}{4}
+\gamma_{12}(j)\gamma_{21}(j)}\,.
\end{align}
The projectors $A_\pm(j)$, $A_\pm^2(j)=A_\pm(j)$, $A_+(j)A_-(j)=0$ are given by the
following expression
\begin{equation}
A_\pm(j)=\frac12\left[\mathbb{I}\mp\frac{1}{\Delta(j)} \left(\begin{array}{cc}
\frac{\gamma_{11}(j)-\gamma_{22}(j)}{2}
&\gamma_{12}(j)\\
\gamma_{21}(j)&\frac
{\gamma_{22}(j)-\gamma_{11}(j)}{2}
 \end{array}\right)\right]\,.
\end{equation}
Let us note that although each term on the r.h.s of (\ref{Lexp})
has  branching points, their sum is an analytic function.
For large $j$, $\gamma_+(j)\simeq \gamma_{11}(j)\equiv \gamma_q(j)$ and
$\gamma_-(j)\simeq \gamma_{22}(j)\equiv \gamma_g(j)$ so we shall refer to $\gamma_{+}(j)(\gamma_-(j))$
as the quark (gluon) anomalous dimension. Let us notice that  due to mixing the quark
(but not gluon) anomalous dimension, $\gamma_+(j)$, has a pole at $j=2$.

The formulae (\ref{Ev-S}), (\ref{ccs}) represent the solution of the
evolution equation
for the quark and gluon GPDs.

Having set $L=1$ in Eq.~(\ref{Ev-S}) one derives a representation analogous to
that in the isovector case, Eq.~(\ref{E0}).
\begin{multline}\label{ES0}
\vec{\mathcal F}^\mu(x,\xi)=\vec{\Delta}^\mu(x,\xi)+\sum_{j=3}^\infty
{\widehat p}_j\left(\frac{x}{|\xi|}\right)\vec{c\,\,}_\xi^{\mu}(j)
+
\sum_{a=\pm}\frac{1}{2 i}\int_{C}
dj\,\cot\pi j\,{\widehat{\mathcal P}_j}^{\,\,a}\left(\frac{ax}{|\xi|}\right)
 \vec{c\,\,}^{\mu,a}_{\xi}(j) ,
\end{multline}
where the integration follows the line $\Re j=1/2$. The function
$\widehat{\mathcal P}_j$ is defined as
\begin{equation}
{\widehat{\mathcal P}_j}^{\,\, \pm}(x)=\left(\begin{array}{cc} {\pm\mathcal P}_j^{(1)}(x)&0\\
0&\xi\, {\mathcal P}_j^{(2)}(x)\end{array}\right)\,,
\end{equation}
where
\begin{equation}\label{PP-m}
{\mathcal P}_j^{(k)}(x)=\theta(x-1)\,(x^2-1)^{k/2} P_{j-1}^{-k}(x)\,.
\end{equation}
The vector $\vec{\Delta}^\mu(x,\xi)$ is given by the following expression
\begin{align}\label{Dk}
\Delta_1^\mu(x,\xi)=&\theta(|\xi|-|x|)\,\frac{x}{|\xi|}\,{\mathcal F}_1^\mu(|\xi|,\xi)\,,\\[2mm]
\Delta_2^\mu(x,\xi)=&\theta(|\xi|-|x|)\left(
{\mathcal F}_2^\mu(|\xi|,\xi)+
\frac{(x^2-\xi^2)}{2|\xi|}\left[\frac{d}{dx}{\mathcal F}_2^\mu\right](|\xi|,\xi)
\right )\,.
\end{align}
The expansion coefficients at  the scale $\mu'$ are given by the same
expressions as in the non-singlet case, Eqn. (\ref{cnew}),
which should now be understood as vector equations.

The anomalous dimension $\gamma_+(j)$ vanishes at $j=3$, so that for $\mu'\to\infty$ the
GPD takes the well known asymptotic form~\cite{Rad98,GPV}
\begin{eqnarray}
{\mathcal F}_1^{as}&=&n_f\frac{x}{\xi^3}\left(1-\frac{x^2}{\xi^2}\right) D\,,\\[2mm]
{\mathcal F}_2^{as}&=&C_F\frac{1}{\xi}\left(1-\frac{x^2}{\xi^2}\right)^2 D\,,
\end{eqnarray}
where
\begin{equation}
D=\frac{15}{4\left(n_f+4C_F\right)}\int_{-1}^1dx\left[x {\mathcal F}_1(x,\xi)+
{\mathcal F}_2(x,\xi)\right]\,.
\end{equation}

Similarly to the case of the isovector GPD one can analyze the small $\xi$ behavior of the
quark and gluon GPDs. In the situation when the GPDs are the smooth functions at the input
scale their form after evolution in the small $\xi$ region is determined by the rightmost
pole of the anomalous dimension $\gamma_+(j)$, which is located at $j=2$.
The corresponding asymptotics have the ``double scaling'' form which results from
evaluation of the integral in Eq.~(\ref{Ev-S}) by the saddle point method.
We again consider the integral corresponding to the term with $a=+$
in Eq.~(\ref{Ev-S}).
The expansion coefficients~(\ref{ccs}) can be expressed as follows
\begin{align}\label{cm}
\vec{c\,}_\xi^{\mu,+}(j)\simeq&
\left(\frac{\xi}{2}\right)^{1-j}\,\vec{m}(j)\,,\\[2mm]
\label{mf}
\vec{m}(j)=&
\frac{\Gamma(j+1/2)}{\sqrt{\pi}\Gamma(j-1)}
\left(\begin{array}{c}{-\mathcal F}_1(j-1)\\
(j-2){\mathcal F}_2(j-2)\end{array}\right)\,,
\end{align}
where
\begin{equation}
\vec{\mathcal F}(s)=\int_0^1 dx x^{s-1}\vec{\mathcal F}^\mu(x,\xi)\,.
\end{equation}
Inserting (\ref{cm}) into Eq.~(\ref{Ev-S}) one finds that  the saddle point is
determined
by the equation
\begin{equation}
1-\gamma_-'(j_*)\,\sigma=0,\ \ \ \ \  \sigma=\frac{\log L}{\log 2/\xi}\,.
\end{equation}
For small $\sigma$ one finds $j=2+\epsilon$ with
\begin{equation}
\epsilon=2\sqrt{\frac{N_c}{b_0}\sigma}\,\left(1+{\mathcal O}(\sigma)\right)\,.
\end{equation}
Then one finds for the integral $\vec{\mathcal I}$ (which corresponds to
the term  $a=+$ in  Eq.~(\ref{Ev-S}) ) in the ``double scaling'' approximation
\begin{align}
\vec{\mathcal I }(x,\xi)
\simeq\left(\frac{\pi^2 N_c}{4b_0}\right)^{1/4}\,\frac{\log^{1/4}L}{\log^{3/4} 2/\xi}\,
e^{4\sqrt{N_c/b_0\log L\log 2/\xi}}
\,\frac{2}{\xi}\,
\widehat q_{2+\epsilon}^{\,\,+}(r)\, A_+(2+\epsilon)\, \vec{m}(2+\epsilon).
\end{align}
Here $r=x/\xi$ and we assume that $\xi>0$. Due to symmetry properties  taking
into account the term $a=-$ results in the replacement
$$
\widehat q_{2+\epsilon}^{\,\,+}(r)\to
\widehat q_{2+\epsilon}^{\,\,+}(r)+\widehat q_{2+\epsilon}^{\,\,-}(-r)$$
in the above expression.

\setcounter{equation}{0}
\section{Summary}\label{summary}
We developed a method for
solving   the leading order evolution equations  for  Generalized Parton Distributions.
The form of the solution,  Eqs.~(\ref{Evp}) and (\ref{Ev-S}),
is fixed completely by the symmetry properties of
the evolution kernels.   We have shown that the GPD should be treated
differently in the ERBL and DGLAP regions, which reflects  both the mathematical
structure of the equation and its physical content.

The Eqs.~(\ref{Evp}) and (\ref{Ev-S})
can be used for both analytical and numerical
studies of the GPD's evolution. It is important to stress here that all quantities
entering these equations are unambiguously defined.
We demonstrated that in the limits $\xi\to 1$ and \mbox{$\xi\to 0$}
the  solutions~(\ref{Evp}) and (\ref{Ev-S}) of the GPD evolution equations
take the form of the solutions of the ERBL and DGLAP evolution equations,
respectively. Furthermore, using the representations~(\ref{Evp}) and (\ref{Ev-S})
one can easily obtain the asymptotic form of the GPDs in
different regimes
(large $L$, small $\xi$, and so on).

It would be quite interesting to apply the developed approach to the analysis of
the analytic structure of the anomalous dimensions of twist-3 operators.
It is well known~\cite{Jar} that the knowledge of the energies of the
reggeon bound  states allows  to extract anomalous dimensions of the operators
at the singular point of in the $j-$plane. This correspondence has
been checked for the
twist two-operators. Recently,  methods for the calculation of
multi-reggeon bound states were developed~\cite{DKM-1,DKKM-2,DVL-1} and the predictions for
the anomalous dimensions of higher twist operators at the singular points
were obtained~\cite{KKM,DVL-2}. However, so far it is not
known how to
solve the problem of the analytical continuation of the anomalous dimensions
of higher twist operators. We hope that our approach can be
generalized
to higher twist operators, at least for the class of the
twist-3 operators for which the
evolution equation are known to be integrable~\cite{BDM,BDKM,DKM,Bint}.

\begin{acknowledgments}
The authors are grateful to V.~Braun, M.~Diehl and D.~M\"uller for useful discussions.
This work was supported in part by the grant 03-01-00837
of the Russian Foundation for Fundamental Research (A.M.),
by the  Helmholtz Association (A.M. and A.S., contract number
VH-NG-004), and the Graduiertenkolleg 841 of DFG (M.K.).
\end{acknowledgments}

\appendix
\app
\section{Auxiliary formulae }
In this Appendix we collected formulae which were useful in our analysis.
We use the definition of the Bessel and Legendre functions of Ref.~\cite{GR}.
It follows from the properties of the Legendre functions  that the functions
$p^{(m)}_{j}(x), q^{(m)}_{j}(x)$ defined in (\ref{pj}),~(\ref{qj}) satisfy the
following relations
\begin{subequations}\label{pqap}
\begin{eqnarray}\label{pap}
\frac{d}{dx}p^{(m)}_{j}(x)&=&p^{(m-1)}_{j}(x)\,,\\[2mm]
\frac{d}{dx}q^{(m)}_{j}(x)&=&q^{(m-1)}_{j}(x)\,.
\label{qap}
\end{eqnarray}
\end{subequations}
The function $q_j^{(m)}(x)$ can be  represented in the form
\begin{align}
q_j^{(m)}(x)=&\frac{e^{i\pi j} {\mathcal Q}_{j}^{(m)}(x+i0)-e^{-i\pi j}
{\mathcal Q}_{j}^{(m)}(x-i0)}{2i\sin\pi j}\,,
\end{align}
where
\begin{align}
{\mathcal Q}_{j}^{(m)}(z)=&(z^2-1)^{m/2}\,Q_{j-1}^{-m}(z)\,.
\end{align}
We give here some integrals involving the functions $p_j^{(m)}(x)$ and $q_j^{(m)}(x)$
\begin{subequations}\label{pq-int}
\begin{eqnarray}
\int_{-1}^1 \frac{dx}{1-x}{p_j^{(m)}(x)}&=&(-1)^{m}2^m\Gamma(m)\frac{\Gamma(j-m)}{\Gamma(j+m)}\,,
\\[2mm]
\int_{-1}^\infty \frac{dx}{x+1}{q_j^{(m)}(x)}&=&
\frac{2^m\Gamma(m)}{2\sin\pi j}\frac{\Gamma(j-m)}{\Gamma(j+m)}
\end{eqnarray}
and
\begin{equation}
\int_{-1}^\infty \frac{dx}{x-1\pm i0}{q_j^{(m)}(x)}=
(-1)^{m}2^m\Gamma(m)\frac{e^{\mp i \pi j}}{2\sin\pi j}\frac{\Gamma(j-m)}{\Gamma(j+m)}\,.
\end{equation}
\end{subequations}
The following integrals involving the Legendre functions
were useful
\begin{subequations} \label{I-PQ}
\begin{eqnarray}
\int_{-1}^1dx P_\nu(x)\,P_m(x)&=&\frac{2}{\pi}\frac{(-1)^{m}\sin
\pi\nu}{(\nu-m)(\nu+m+1)}\,,\\[2mm]
\label{IPQ}
\int_{1}^\infty dx P_\nu(x)\,Q_\lambda(x)&=&\frac{1}{(\nu-\lambda)(\nu+\lambda+1)}\,,
\end{eqnarray}
\end{subequations}
where $m$ is integer.
Next, we remind the following relation
\begin{equation}\label{PQ}
P_{\nu}(z)=\frac1\pi \tan\pi \nu\left[Q_{\nu}(z)-Q_{-\nu-1}(z)\right]
\end{equation}
between the Legendre functions of the first and second kind.
Next, for $z\to 1$, the Legendre function $Q_\nu(z)$ has the following behavior
\begin{equation}\label{Qz}
Q_\nu(z)=-\frac12 
\log(z-1)+{\mathcal O}(1).
\end{equation}
It follows from the Eqs.~(\ref{IPQ}) and (\ref{PQ}) that
\begin{equation}
\int_1^\infty dx P_{-1/2+i\rho}(x)P_{-1/2+i\rho'}(x)=
\rho^{-1}\coth\pi\rho\left[\delta(\rho-\rho')+\delta(\rho+\rho')
\right].
\end{equation}
The Fourier transform of the function $\Psi^\xi_j(z)$~(see
Eq.~(\ref{psij})) is
\begin{equation}
\frac{1}{2\pi}\int_{-\infty}^\infty \frac{dz}{z^m} e^{-ixz} \Psi^\xi_j(z+i0)=
\frac{1}{i^m} e^{-i\pi/4}
|\xi|^{m-1/2}\sqrt{2} \pi^{-3/2}\sin\pi j \,q_j^{(m)}\left(\frac{x}{|\xi|}\right)\,,
\end{equation}
where $\Re j>m$.
We also remind that
\begin{equation}\label{P1i}
\int_1^\infty dx x^{-s} P_\nu(x)=2^{s-2}\frac{\Gamma\left(\frac{s+\nu}{2}\right)
\Gamma\left(\frac{s-\nu-1}{2}\right)}{\pi^{1/2}\Gamma(s)}\,,
\end{equation}
\begin{equation}
\int_0^1 dx x^{-s} P_\nu(x)=2^{s-1}
\frac{\pi^{1/2}\Gamma(1-s)}{\Gamma\left(\frac{2-s-\nu}{2}\right)
\Gamma\left(\frac{3-s+\nu}{2}\right)}\,.
\end{equation}



\end{document}